\documentclass[runningheads]{llncs}

\usepackage[utf8]{inputenc}

\usepackage{amsmath}
\usepackage{latexsym}
\usepackage{xspace}
\usepackage{enumitem}
\usepackage{etoolbox}
\usepackage{microtype}

\usepackage{algorithmicx}
\usepackage{algorithm}
\usepackage{algpseudocode}

\newtheorem{principle}{Principle}

\newtheorem{observation}{Observation}

\newcommand{\To}{\Rightarrow}
\newcommand{\defeater}{\leadsto}

\newcommand{\PROP}{\ensuremath{\mathrm{PROP}}\xspace}

\newcommand{\LIT}{\ensuremath{\mathrm{Lit}}\xspace}
\newcommand{\MODLIT}{\ensuremath{\mathrm{DLit}}\xspace}

\newcommand{\FACTS}{\ensuremath{F}\xspace}
\newcommand{\LAB}{\ensuremath{\mathrm{Lab}}\xspace}
\newcommand{\non}{\ensuremath{\mathord{\sim}}}
\newcommand{\seq}[2][n]{#2_{1},\dots,#2_{#1}}
\newcommand{\set}[2][]{\ifblank{#1}{\left\{#2\right\}}{#1\{#2#1\}}}

\let\arrow\hookrightarrow

\newcommand{\OBL}{\ensuremath{\mathsf{O}}\xspace}
\newcommand{\PERM}{\ensuremath{\mathsf{P}}\xspace}

\newcommand{\reduct}{\mathit{red}}

\title{Avoiding Pragmatic Oddity:\break A Bottom-up Defeasible Deontic Logic}
\titlerunning{Avoiding Pragmatic Oddity}
\author{Guido Governatori, Silvano Colombo Tosatto and Antonino Rotolo}
\institute{}

\begin{document}
\maketitle

\begin{abstract}
This paper presents an extension of Defeasible Deontic Logic to deal with the Pragmatic 
Oddity problem. The logic applies three general principles: (1) the Pragmatic Oddity problem must be solved within a general logical treatment of CTD reasoning; (2) non-monotonic methods must be adopted to handle CTD reasoning; (3) logical models of CTD reasoning must be computationally feasible and, if possible, efficient. The proposed extension of Defeasible Deontic Logic elaborates a preliminary version of the model proposed by Governatori and Rotolo \cite{Governatori:2019:A-Computational}. The previous solution was based on particular characteristics of the (constructive, top-down) proof theory
of the logic. However, that method introduces some degree of non-determinism. To avoid the problem, we provide a bottom-up characterisation of the logic. The new characterisation offers insights for the efficient implementation of the logic and allows us to establish the computational complexity of the problem.
\end{abstract}

\section{Introduction}
\label{sec:intro}

A key difference between norms and other constraints is that, typically,
norms can be violated. Moreover, normative systems (especially the legal ones) contain 
provisions about norms that become effective when violations occur. 
Since the seminal work by Chisholm \cite{chisholm1963contrary} the obligations in
force triggered by violations have been dubbed contrary-to-duty 
obligations (CTDs). The treatment of CTDs has proven problematic for
formal (logical) representations of normative systems. Accordingly, CTDs
are the source of many paradoxes and problems and also the driver for criticising Standard Deontic Logic (SDL) and for the development of many new deontic formalisms (see \cite{carmo2002deontic,handbook}). 

One well-known problem of CTDs is the so-called Pragmatic Oddity  
paradox, which was introduced by Prakken and Sergot 
\cite{PrakkenSergot96} and is illustrated by the following example.

\begin{example}\label{ex:PrakkenSergot}
\begin{align}
\text{There should be no dog.} \qquad  & \OBL \neg d \label{eq:OBL_NoDog}\\
\text{If there is a dog, then there ought to be a warning sign.} \qquad & d\to \OBL s \label{eq:OBL_Sign}\\
\text{There is a dog.} \qquad & d \label{eq:Dog}
\end{align}
\end{example}

In SDL, we have both $\OBL \neg d$ and $\OBL s$. However, according to Prakken and Sergot, 
\begin{quote}
``Surely, it is strange to say that in all ideal worlds there is no dog and also a warning sign that there is no dog. [\dots]
This oddity---we might call it a `pragmatic oddity'---seems to be absent from the natural language version, which means that the SDL representation is not fully adequate.'' \cite[pp. 96, 95]{PrakkenSergot96}
\end{quote}

The oddity of Example \ref{ex:PrakkenSergot}, its counter-intuitiveness, seems to depend on the fact that the two obligations  $\OBL \neg d$ and $\OBL s$ are in force at the same time: when you fail to have no dog, you are obliged to have no dog and obliged to hang a warning sign. The solutions proposed by Prakken and Sergot \cite{PrakkenSergot96} consist of representing (\ref{eq:OBL_Sign}) as $d\To \OBL_d s$, where $\To$ is a suitable conditional operator. The sentence $\OBL_d s$ means that ``there is a secondary obligation that $s$, presupposing the sub-ideal context $d$''. The problem,  for Prakken and Sergot, is avoided because $\OBL_d s$ does not imply $\OBL s$: ``primary and CTD obligations are obligations of a different kind: a CTD obligation pertains to, or presupposes, a certain context in which a primary obligation is already violated'' \cite[p. 91]{PrakkenSergot96}. 

Prakken and Sergot's analysis is thus based on two basic principles:
\begin{principle}\label{pr:CTD}
The Pragmatic Oddity problem must be solved within a general logical treatment of CTD reasoning.
\end{principle}
\begin{principle}\label{pr:classes}
Primary obligations and CTD obligations are of a different kind.
\end{principle}

In fact, most of the work on Pragmatic Oddity (in addition to \cite{PrakkenSergot96}, see, among others \cite{carmo2002deontic})\footnote{For excellent overviews of the literature, see \cite{carmo2002deontic,handbook:goble,parent2017pragmatic}.}
focuses on the issue of how to distinguish the mechanisms leading to the 
derivation of the two individual obligations, and create different classes of 
obligations insofar as they express different ideality levels. 
One solution is to prevent the conjunction when the obligations are from different classes.
Accordingly, if the problem is to avoid having a conjunctive obligation in 
force when the individual obligations are in force themselves, the simplest
way is to have a deontic logic that does not support the aggregation 
axiom:
\[
  (\OBL a\wedge \OBL b) \rightarrow \OBL(a \wedge b)
\] 

This solution, among other things, was discussed by \cite{Goble05}: adopting a non-normal deontic logic (i.e., weaker than $\mathbf{K}$), each obligation semantically corresponds to a distinct norm that selects a set of ideal worlds (see \cite{Multiplex}) and aggregation cannot be allowed.

However, as suggested by \cite{Horty93deonticlogic} and also recalled by \cite{Goble05}, some restricted forms of agglomeration should be accepted: several examples seem to hold if CTDs are not considered.

Therefore, a more amenable option, as suggested by Parent and van der 
Torre~\cite{parent2014sing,parent2017pragmatic}, is to admit aggregation for 
obligations that are independent of the violation of the other obligations. We agree with them. Indeed, in our view, what is odd is not that the two obligations are in force 
at the same time, but that if one admits forming a conjunctive 
obligation from the two individual obligations, then we get an 
obligation that is impossible to comply with. 

Based on the intuition above, Governatori and Rotolo~\cite{Governatori:2019:A-Computational} proposed, in a preliminary work, an extension of Defeasible Deontic Logic~\cite{jpl:permission} to handle Pragmatic 
Oddity. Their solution was based on the constructive proof theory of the logic 
with specific proof conditions: more specifically, to admit the derivation of 
$\OBL(a \wedge b)$ requires that $\OBL a$ and $\OBL b$ are already provable, 
and $\neg a$ does not appear in the derivation of $\OBL b$ (similarly for 
$\OBL a$ and $\neg b$).

The extension mentioned above of Defeasible Deontic Logic was based on Principle \ref{pr:CTD} by using the new non-classical operator $\otimes$: the reading of an expression like $a\otimes b \otimes c$ is that $a$ is primarily obligatory, but if this obligation is violated, the secondary obligation is $b$, and, if the secondary (CTD) obligation $b$ is violated as well, then $c$ is obligatory (see \cite{jpl:permission}). This approach falls within a proof-theoretic line of inquiry about CTDs, which clearly distinguishes in the language and the logic structures representing norms from those representing obligations \cite{makinson00input,VanDerTorre:2001,Makinson:1999}. 

We also complied with another principle:

\begin{principle}\label{pr:NM}
Non-monotonic methods must be adopted to handle CTD reasoning. 
\end{principle}

This principle was notably defended by  Horty \cite{Horty93deonticlogic} and van der Torre and Tan \cite{vanderTorre1997}, even though, according to Parent and van der Torre \cite{parent2017pragmatic} it seemed not directly involved in the Pragmatic Oddity problem. Since we stick to Principle \ref{pr:CTD}, we also adopt Principle \ref{pr:NM}. However, we will see in Section \ref{sec:scenarios}  that the idea of defeasibility plays a role in some scenarios of Pragmatic Oddity, too. 

Finally, our concern was computational since CTDs are so pervasive in normative reasoning (for example, in the law):
\begin{principle}\label{pr:complexity}
Logical models of CTD reasoning must be computationally feasible and, possibly, efficient. 
\end{principle}

The solution we proposed in \cite{Governatori:2019:A-Computational} was
based on particular characteristics of the (constructive, top-down) proof theory
of the logic. However, that method introduces some degree of non-determinism,
insofar as the solution requires the existence of a proof satisfying certain
conditions, and alternative proofs are possible. 

The contribution of this paper is to present a new logical framework for the Pragmatic Oddity problem, which 
\begin{itemize}
\item revises the solution we advanced in \cite{Governatori:2019:A-Computational}, 
\item complies with the same general principles we set in \cite{Governatori:2019:A-Computational},
\item provides a bottom-up characterisation of the logic that avoids the problem
with the top-down solution,
\item studies the complexity of the problem (the
resulting logic is computationally feasible, i.e., polynomial in the size of the
input theory); and offers insights for the efficient implementation of the logic.
\end{itemize}

The layout of the article is as follows. Section~\ref{sec:deflogic} offers a 
high-level introduction to Defeasible Deontic Logic, while
Section \ref{sec:ddl-po} presents the technical details of the Defeasible Deontic Logic used in the paper, a logic equipped with the $\otimes$-operator to identify pragmatic oddity instances. Section \ref{sec:scenarios} discusses some examples and scenarios of pragmatic oddity. Section \ref{sec:bottom-up} provides a bottom-up characterisation of the logic. Section \ref{sec:complexity} studies the computational complexity of 
the problem of computing whether a conjunctive obligation is derivable 
from a given defeasible theory. The paper ends with some brief conclusions.

\section{A Gentle Overview of Defeasible Deontic Logic}
\label{sec:deflogic}

This section provides a gentle overview of Defeasible Deontic
Logic and how to use it for normative reasoning (see also Section \ref{sec:ddl-po}). For more detailed
presentations of the logic and its uses to model different aspects 
of normative reasoning, we refer the readers to \cite{jpl:permission,rr2018-school,logic-legislation,handbook:deontic}.

Defeasible Deontic Logic \cite{jpl:permission} is a sceptical computationally
 oriented rule-based formalism designed for the representation of norms. 
 The logic extends Defeasible Logic~\cite{tocl} with deontic operators to 
 model obligations and (different types of) permissions and provides an 
 integration with the logic of violation developed by Governatori and Rotolo~\cite{ajl:ctd}. The 
 resulting formalism offers features for the natural and efficient 
 representation of exceptions, constitutive and prescriptive rules, and
compensatory norms. The logic is based on a constructive proof theory that 
allows for full traceability of the conclusions and flexibility to handle and
combine different facets of non-monotonic reasoning. 

Knowledge in Defeasible Logic is structured in three components:
\begin{itemize}
\item A set of facts (corresponding to indisputable statements represented as 
  literals, where a literal is either an atomic proposition or its negation).
\item A set of rules. A rule establishes a connection between a set of premises 
  and a conclusion.  In particular, for reasoning with norms, it is reasonable 
  to assume that a rule provides the formal representation of a norm (though, 
  it is possible to have norms that are represented by a set of rules). Accordingly,
  the premises encode the conditions under which the norm is applicable, and the
  conclusion is the normative effect of the norm.
\item A preference relation over the rules.  The preference relation just gives
  the relative strength of rules. It is used in contexts where two rules with
  opposite conclusions fire simultaneously to determine that one rule 
    overrides the other in that context.
\end{itemize}
The rules establish a relationship between a set of premises (the antecedent) of 
a rule and a conclusion. We can classify rules based on (1) the strength of the
relationship and (2) the type of relationship, more precisely, the type (or mode) of 
conclusion or effect a rule produces.  Accordingly, a rule is an expression 
\begin{equation}
	\seq{a}\hookrightarrow_\Box c
\end{equation}
where $\seq{a}$ is the antecedent, $c$ is the conclusion, $\hookrightarrow$ indicates 
the strength and $\Box$ the mode. For the strength Defeasible Logic provides three 
kinds of rules: \emph{strict rules} (represented by $\to$), \emph{defeasible rules} 
(represented by $\To$), and \emph{defeaters} (represented by $\leadsto$).  A strict 
rule is a rule in the classical sense; every time the antecedent holds, so does the
conclusion. On the other hand, a defeasible rule can produce its effect (or
conclusion) when it is applicable and when there are no (applicable) rules for 
the opposite or such rules are defeated (by stronger rules). Finally, defeaters 
are rules that do not directly produce a conclusion but prevent the opposite 
conclusion from holding.  

For the type or mode of the conclusion, we distinguish between \emph{constitutive 
rules} and \emph{normative rules}.  Constitutive rules are used to define terms 
as defined in the normative systems the rules are meant to formalise. Therefore, 
constitutive rules specify the institutional facts or statements that 
hold in a given situation. Thus, for example, the constitutive rule 
\[
 person, age<18y \To minor
\]    
establishing  the institutional fact that minors are persons whose age is less 
than 18 years (for the notation we drop the $\Box$ for constitutive rules). On 
the contrary, a normative rule determines the conditions under which the conclusion 
is in force as an obligation or permission (one of the two modal operators of 
Defeasible Deontic Logic). Consider, for instance, the following two normative rules:
\begin{gather}
r_1\colon vehicle, redLight \To_\OBL stop \\
r_2\colon emergency, redLight \To_\PERM \neg stop
\end{gather}
The first, $r_1$ is a prescriptive rule (indicated by the obligation $\OBL$ modality)
prescribing the obligation to stop for vehicles approaching a set of red traffic lights. 
Thus, when the conditions set in the antecedent hold ($vehicle$ and $redLight$), the 
rule allows us to conclude the obligation to stop $\OBL stop$ is in force. $r_2$ is a
\emph{permissive} rule derogating or establishing an exception to $r_1$ for emergency 
vehicles. When its antecedent holds, we can conclude that it is permitted not to stop 
($\PERM\neg stop$). The two rules conflict with each other, and we can use 
the superiority relation to state that $r_2$ overrides $r_1$, namely $r_2 > r_1$.  

As we mentioned, a characteristic of normative reasoning is its ability to deal with 
violations and conditions triggered by them. To this end, Defeasible Deontic Logic 
extends the language with a compensation operator $\otimes$ to form 
expressions like 
\[
 c_1 \otimes c_2 \otimes \cdots\otimes c_n
\]
called \emph{compensation chains}.  Compensation chains are only allowed as
the conclusion of prescriptive rules (and thus asserting that obligations are in 
force). Their meaning 
as proposed by Governatori and Rotolo \cite{ajl:ctd} and further discussed by  Governatori \cite{icail2015:thou},
is that $\OBL c_1$ is the primary obligation, and when violated (i.e., 
$\neg c_1$ holds), then $\OBL c_2$ is in force, and it compensates for the 
violation of the obligation of $c_1$. Moreover, when $\OBL c_2$ is violated, 
then $\OBL c_3$ is in force, and so on until we reach the end of the chain when 
a violation of the last element is a non-compensable violation where the norm
corresponding to the rule in which the chain appears is not complied with.

Defeasible Logic is a constructive logic. Hence, the kernel of the logic is 
its proof theory, and for every conclusion we draw from 
a defeasible theory we can provide a proof for it, giving the steps 
used to reach the conclusion.  At the same time, the derivation gives a (formal)
explanation or justification of the conclusion. Furthermore, the logic
distinguishes between \emph{positive} and \emph{negative} conclusion, the strength
of a conclusion and its mode. This is achieved by labelling each step in a 
derivation with a proof tag. A derivation is a (finite) sequence of (tagged) 
formulas, each obtained from the previous ones using inference conditions. 
The inference conditions are formulated as proof conditions mandating the conditions 
that the previous steps in a derivation have to satisfy to append a 
new conclusion as the next step of a derivation.   We adopt the following notation 
for proof tags: $+$ and $-$ indicate whether we have a positive or negative 
conclusion, $\Delta$ and $\partial$ denote, respectively,  a definite or a 
defeasible conclusion, and they are subscripted by the modal (deontic) operator 
describing the mode of the conclusion. For example, the meaning of the tagged 
literal $-\Delta_Cp$ is that we definitely refute $p$ as an institutional 
fact\footnote{Similarly to the notation used for rules we drop the subscript for 
constitutive conclusions.}. This means that we explored all possible ways to prove 
$p$ using constitutive rules and facts, and we failed to derive it. On the other hand, 
$+\partial_\OBL\neg p$ means that we have a defeasible derivation for $\neg p$, where 
the rule used to conclude is a prescriptive rule. Finally, we say that 
$\Box p$ is provable if we have a positive derivation for $p$ with mode $\Box$. 
Accordingly, $\OBL p$ holds if we derive $+\partial_\OBL p$ (or the stronger 
$+\Delta_\OBL p$). 
  
Defeasible derivations have a three-phase argumentation-like
structure. To
show that $+\partial_\Box p$ is provable at step $n$ of a derivation we have
to:\footnote{Here we concentrate on proper defeasible derivations.}
\begin{enumerate}
  \item give an argument for $p$ (where the last rule is a rule for $\Box$);
  \item consider all counterarguments for $p$; and 
  \item rebut each counterargument by either: 
  \begin{enumerate}
    \item showing that the counterargument is not valid;
    \item providing a valid argument for $p$ defeating the counterargument.
  \end{enumerate}
\end{enumerate}
In this context, in the first phase, an argument is simply a strict or
defeasible rule for the conclusion we want to prove, where all the elements
are at least defeasibly provable. In the second phase, we consider all rules
for the opposite or complement of the conclusion to be proved.
Here, an argument (counterargument) is not valid if the argument is not
supported. Here ``supported'' means that all the
elements of the body are at least defeasibly provable. 

Finally, to defeasibly refute a literal, we have to show that either, the
opposite is at least defeasibly provable, or an exhaustive search
for a constructive proof for the literal fails (i.e., there are no rules for such
a conclusion, or all rules are either `invalid' arguments or they are not
stronger than valid arguments for the opposite).

\section{A Defeasible Deontic Logic for Pragmatic Oddity}
\label{sec:ddl-po}

In this section, we present a variant of Defeasible Deontic Logic
designed to deal with the issue of Pragmatic Oddity. More specifically, we
show how the proof theory can be used to propose a simple and
(arguably) elegant treatment of the problem at hand.  

We restrict ourselves to the fragment of Defeasible Deontic
Logic that excludes permission and permissive rules since they do not affect 
the way we prevent Pragmatic Oddity from occurring: Definitions~\ref{def:+pand} and 
\ref{def:-pand}, the definitions that describe the mechanisms we adopt for
a solution to Pragmatic Oddity are independent of any issue related to
permission. In addition, for the sake of simplicity and to better focus on the 
non-monotonic aspects that the logic offers, we use only defeasible rules and 
defeaters. However, the definitions can be used directly in the full version 
of the logic. Accordingly, we consider a logic whose language is defined as 
follows.

\begin{definition}
	Let \PROP be a set of propositional atoms and \OBL the
modal operator for obligation.
\begin{itemize}
	\item The set $\LIT=\PROP\cup \{ \neg p\,|\,p\in\PROP\}$ is the set of \emph{literals}.
	\item The \emph{complement} of a literal $q$ is denoted by $\non q$; if $q$ is a positive literal $p$, then $\non q$ is $\neg p$, and if $q$ is a negative literal $\neg p$, then $\non q$ is $p$.
	\item The set of \emph{deontic literals} is $\MODLIT=\{\OBL l, \neg \OBL l\,|\,l\in \LIT\}$.
	\item If $c_1,\dots,c_n\in\LIT$, then $\OBL(c_1\wedge\dots\wedge c_n)$ is a \emph{conjunctive obligation}.
\end{itemize}
\end{definition}
In the rest of the paper, when relevant to the discussion, we will refer 
to elements of $\LIT$ as plain literals, and often we will use the 
unmodified term `literal' to indicate either a plain literal or a 
deontic literal. 
 
We formally introduce the compensation operator $\otimes$. 
This operator is used to build chains of compensation called $\otimes$-expressions. The formation rules for well-formed
$\otimes$-expressions are:
\begin{enumerate}
  \item every literal $l \in \LIT$ is an $\otimes$-expression;
  \item if 
    $c_1, \dots, c_k \in \LIT$, then $c_1 \otimes \dots \otimes c_k$ 
    is an $\otimes$-expression;
  \item nothing else is an $\otimes$-expression.
\end{enumerate}
 
Given an $\otimes$-expression $A$, the \emph{length} of $A$ is the
number of literals in it. Given an $\otimes$-expression $A\otimes b\otimes C$
(where $A$ and $C$ can be empty), the \emph{index} of $b$ is the length of $A
\otimes b$. We also say that $b$ appears at index $n$ in $A \otimes b$ if the
length of $A \otimes b$ is $n$.

\begin{definition}
	Let \LAB be a set of arbitrary labels. Every rule is of the type
\begin{displaymath}
r\colon A(r) \arrow C(r)
\end{displaymath}
where
\begin{enumerate}
\item $r \in \LAB$ is the name of the rule; 
\item $A(r) = \set{\seq{a}}$, the \emph{antecedent} (or \emph{body}) of the
rule, is the set of the premises of the rule (alternatively, it can be
understood as the conjunction of all the elements in it). Each $a_i$ is either
a literal, a deontic literal or a conjunctive obligation;
\item $\arrow\in\{\Rightarrow,\Rightarrow_{\OBL}, \leadsto, \leadsto_{\OBL}\}$ denotes the type of the rule.
If $\arrow$ is $\Rightarrow$, the rule is a \emph{defeasible rule}, while
if $\arrow$ is $\leadsto$, the rule is a \emph{defeater}. Rules without the
subscript \OBL are constitutive rules, while rules with such a subscript are
prescriptive rules. 
\item $C(r)$ is the \emph{consequent} (or \emph{head}) of the rule. It is a 
  single literal for defeaters and constitutive rules, and an $\otimes$-expression
  for prescriptive defeasible rules.
\end{enumerate}
\end{definition}
Recall that prescriptive rules are used to derive obligations.

Given a set of rules $R$, we use the following abbreviations
for specific subsets of rules:
\begin{itemize}
\item $R_{d}$ denotes the set of defeasible rules in the set $R$;
\item $R[q,n]$ is the set of rules where $q$ appears at index $n$ in the
consequent.\footnote{Strictly speaking, the notion of index is defined for $\otimes
$-expressions and not for literals; however, according to the construction 
rules for $\otimes$-expressions a plain literal is an $\otimes$-expression.} 
The set of rules where $q$ appears at any index $n$ is denoted by $R[q]$;
\item $R^{\OBL}$ denotes the set of prescriptive rules in $R$, i.e., the set of rules with \OBL as their subscript;
\item $R^C$ denotes the set of constitutive rules in $R$, i.e., $R\setminus R^\OBL$.
\end{itemize}
The above notations can be combined. Thus, for example, $R^{\OBL}_d[q,n]$ stands
for the set of defeasible prescriptive rules such that $q$ appears at index $n$ 
in the consequent of the rule.  

\begin{example}\label{ex:rules}
Let us consider the following set of rules $R$:
\begin{align*}
 r_1\colon f_1 &\Rightarrow_{\OBL} a \otimes b &
 r_2\colon f_2,g_2 &\Rightarrow_{\OBL} b \otimes c &
 r_3\colon f_3 &\Rightarrow \neg a\\
 r_4\colon d &\leadsto_{\OBL} \neg a	 &
 r_5\colon \neg\OBL a&\Rightarrow \neg b &
 r_6\colon \OBL a, \OBL b &\Rightarrow_{\OBL} \neg c\\
 && 
 r_7\colon f_7&\Rightarrow d
\end{align*}
The set of prescriptive rules $R^\OBL$ is  $\set{r_1, r_2,r_4, r_6}$; accordingly, 
the set of constitutive rules $R^C=\set{r_3, r_5, r_7}$. Moreover, the set of
prescriptive defeasible rules $R^{\OBL}_d=\set{r_1,r_2,r_6}$. The set of rules for 
$\neg a$, $R[\neg a]$ is $\set{r_3, r_4}$; notice that this set contains a
prescriptive and a constitutive rule; the corresponding set of defeasible
constitutive rules $R^{C}_d[\neg a]=\set{r_3}$.  When we consider the index where 
a literal appears we have the following sets: $R[b,1]=\set{r_2}$, $R[b,2]=\set{r_1}$
and $R[b]=\set{r_1, r_2}$.
\end{example}

\begin{definition}
A \emph{Defeasible Theory} is a structure $D = (F,R,>)$ where $F$, the set of
facts, is a set of (plain) literals, $R$ is a set of rules, and $>$,
the superiority relation, is a binary relation over $R$.
\end{definition}
A theory corresponds to a normative system, i.e., a set of norms, where every 
norm is modelled by some rules; accordingly, we do not admit deontic literals in the 
set of facts; obligations are determined by norms, and hence, in our framework by 
prescriptive rules. If both rules fire, the superiority relation is used for conflicting
rules, i.e., rules whose conclusions are complementary literals. We do not restrict the superiority
relation: it just determines the relative strength between
two rules.

\begin{definition}
A \emph{proof} (or derivation) $P$ in a defeasible theory $D$ is a linear sequence $P(1)\dots
P(z)$ satisfying the
proof conditions given in Definitions~\ref{def:+p}--\ref{def:-pand}, and each $P(i)$, $1\leq i \leq z$, is a \emph{tagged expression}, i.e., an expression of one of the forms: $+\partial q$, $-\partial q$, 
$+\partial_{\OBL} q$, $-\partial_{\OBL}q$, $+\partial_{\OBL}c_1\wedge\cdots\wedge c_m$ and $-\partial_{\OBL}c_1\wedge\cdots\wedge c_m$.
\end{definition}
The tagged literal $+\partial q$ means that $q$ is \emph{defeasibly provable} 
as an institutional statement, or in other terms, that $q$ holds in the 
normative system encoded by the theory. The tagged literal $-\partial q$ 
means that $q$ is \emph{defeasibly refuted} by the normative system.
Similarly, the tagged literal $+\partial_{\OBL} q$ means that $q$ is
\emph{defeasibly provable} in $D$ as an obligation or that $\OBL p$ is defeasibly provable. In contrast,
$-\partial_{\OBL} q$ means that $q$ is \emph{defeasibly refuted} as an obligation, 
thus $\OBL p$ cannot be proved.
For $+\partial_{\OBL}c_1\wedge\cdots\wedge c_m$ the meaning is that 
the conjunctive obligation $\OBL(c_1\wedge\cdots\wedge c_m)$ is defeasibly 
derivable; and that a conjunctive obligation $\OBL(c_1\wedge\cdots\wedge c_m)$ 
is defeasibly refuted corresponds to $-\partial_{\OBL}c_1\wedge\cdots\wedge c_m$. 
The initial part of length $i$ of a proof $P$ is denoted by $P(1..i)$.

Defining when a rule is applicable or discarded is essential to characterise
the notion of provability for constitutive rules and then for obligations. A
rule is \emph{applicable} for a literal $q$ if $q$ occurs in the head of the
rule and all elements in the antecedent have been defeasibly proved (eventually
with the appropriate modalities). On the other hand, a rule is \emph{discarded} 
if at least one of the modal literals in the antecedent has not been proved. 
However, as literal $q$ might not
appear as the first element in an $\otimes$-expression in the head of the
rule, some additional conditions on the consequent of rules must be satisfied.
Accordingly, we first define the case for a constitutive rule (body-applicable)
before moving to the condition for prescriptive rules with $\otimes$-expressions 
(Definition~\ref{definition:APPL+pO}, where a literal is applicable if the 
previous element is provable as an obligation but violated, meaning that its 
complement is derivable. 

\begin{definition}\label{definition:BodyApplicable} Given a proof $P$,
	a rule $r \in R$ is \emph{body-applicable} at step $P(n+1)$ iff for all $a_i \in A(r)$:
\begin{enumerate}
  \item if $a_i = \OBL l$ then $+\partial_{\OBL} l \in P(1..n)$;
  \item if $a_i = \neg \OBL l$ then $-\partial_{\OBL} l \in P(1..n)$;
  \item if $a_i = \OBL(c_1\wedge\cdots\wedge c_m)$ then $+\partial_{\OBL}
     c_1\wedge\cdots\wedge c_m\in P(1..n)$;
  \item if $a_i = l \in \LIT$ then $+\partial l\in P(1..n)$.
\end{enumerate}
A rule $r \in R$ is \emph{body-discarded} at step $P(n+1)$ iff $\exists a_i \in A(r)$ such that
\begin{enumerate}
  \item if $a_i = \OBL l$ then $-\partial_{\OBL} l \in P(1..n)$;
  \item if $a_i = \neg \OBL l$ then $+\partial_{\OBL} l \in P(1..n)$;
  \item if $a_i = \OBL(c_1\wedge\cdots\wedge c_m)$ then $-\partial_{\OBL}
     c_1\wedge\cdots\wedge c_m\in P(1..n)$;
  \item if $a_i = l \in \LIT$ then $-\partial l\in P(1..n)$.
\end{enumerate}
\end{definition}

\begin{definition}\label{definition:APPL+pO}
  Given a proof $P$,
a rule $r \in R^\OBL[q,j]$ such that $C(r) = c_{1} \otimes \cdots \otimes c_{m}$ is \emph{applicable} for literal $q$ at
index $j$ at step $P(n+1)$ (or, simply, applicable for $q$), with $1 \leq j < m$, in the condition for $\pm\partial_{\OBL}$ iff
\begin{enumerate}[nosep]
  \item $r$ is body-applicable at step $P(n+1)$; and
  \item for all $c_k \in C(r)$, $1 \leq k < j$, $+\partial_{\OBL}c_k\in P(1..n)$ and $+\partial\non c_k \in P(1..n)$.
\end{enumerate}
\end{definition}
\noindent
Condition (1) represents the requirements on the antecedent stated in
Definition~\ref{definition:BodyApplicable}; condition (2) on the head of the
rule states that each element $c_k$ before $q$ has been derived as an
obligation and a violation of such obligation has occurred.

\begin{definition}\label{definition:DISC+pO+pP}
  Given a proof $P$,
a rule $r \in R^{\OBL}[q,j]$ such that $C(r) = c_{1} \otimes \cdots \otimes c_{m}$ is \emph{discarded} for literal $q$ at
index $j$ at step $P(n+1)$ (or, simply, discarded for $q$), with $1 \leq j \leq m$, in the condition for $\pm\partial_{\OBL}$ iff
\begin{enumerate}[nosep]
  \item $r$ is body-discarded at step $P(n+1)$; or 
  \item there exists $c_k \in C(r)$, $1 \leq k < l$, such that either $-\partial_{\OBL} c_k\in P(1..n)$ or $-\partial\non c_k \in P(1..n)$. 
\end{enumerate}
\end{definition}
\noindent
In this case, condition (2) ensures that an obligation before $q$
in the chain is not in force or has already been fulfilled (thus, no
reparation is required).

We now introduce the proof conditions for $\pm\partial$ and $\pm\partial_{\OBL}$:

\begin{definition}\label{def:+p}
The proof condition of \emph{defeasible provability for an institutional statement} is

\smallskip \noindent
\begin{minipage}{\textwidth}
\begin{tabbing}
  $+\partial$: If $P(n+1)=+\partial q$ then\\
  (1) \= $q \in \FACTS$ or\\
  \> (2.1) $\non q \not\in \FACTS$ and\\ 
  \> (2.2) $\exists r\in R_d[q]$ such that $r$ is applicable for $q$, and \\
  \> (2.3) \= $\forall s\in R[\non q]$, either \\
      \> \> (2.3.1) $s$ is discarded for $\non q$, or\\
      \> \> (2.3.2) $\exists t\in R[q]$ such that $t$ is applicable for $q$ and $t> s$.
\end{tabbing}
\end{minipage}
\end{definition}
\noindent
As usual, we use the strong negation to
define the proof condition for $-\partial$

\begin{definition}\label{def:-p}
The proof condition of \emph{defeasible refutability for an institutional 
statement} is

\smallskip \noindent
\begin{minipage}{\textwidth}
\begin{tabbing}
  $-\partial$: If $P(n+1)=-\partial q$ then\\
  (1) \= $q \notin \FACTS$ and\\
  \> (2.1) $\non q \in \FACTS$ or\\ 
  \> (2.2) $\forall r\in R_d[q]$: either $r$ is discarded for $q$, or \\
  \> (2.3) \= $\exists s\in R[\non q]$, such that \\
      \> \> (2.3.1) $s$ is applicable for $\non q$, and\\
      \> \> (2.3.2) $\forall t\in R[q]$ either $t$ is discarded for $q$ or  $t\not> s$.
\end{tabbing}
\end{minipage}
\end{definition}
\noindent
The proof conditions for $\pm\partial$ are the standard conditions in Defeasible 
Logic, see \cite{tocl} for the full explanations. 

\begin{definition}\label{def:+pO}

The proof condition of \emph{defeasible provability for obligation} is

\smallskip \noindent
\begin{minipage}{\textwidth}
\begin{tabbing}
  $+\partial_{\OBL}$: If $P(n+1)=+\partial_{\OBL} q$ then\\
  (1) $\exists r\in R_d^{\OBL}[q,i]$ such that $r$ is applicable for $q$, and \\
  (2) \= $\forall s\in R^\OBL[\non q, j]$, either \\
     \> (2.1) $s$ is discarded for $\non q$, or\\
     \> (2.2) $\exists t\in R^\OBL[q,k]$ such that $t$ is applicable for $q$ and $t> s$.
\end{tabbing}
\end{minipage}
\end{definition}
\noindent
To show that $q$ is defeasibly provable as an obligation, one must show 
that: the following two conditions must hold:  (1) there must be a rule
introducing the obligation for $q$ which can apply; (2) every rule $s$ for
$\non q$ is either discarded or defeated by a stronger rule for $q$.  Observe that, 
since we do not admit deontic literals in $\FACTS$, we do not need the 
equivalent of conditions (1) and (2.1) for institutional statements to ensure 
that the logic is consistent.

The strong negation of Definition~\ref{def:+pO} gives the negative proof
condition for obligation.

\begin{definition}\label{def:-pO}

The proof condition of \emph{defeasible refutability for obligation} is

\smallskip \noindent
\begin{minipage}{.95\textwidth}
\begin{tabbing}
  $-\partial_{\OBL}$: If $P(n+1)=-\partial_{\OBL} q$ then\\
  (1) $\forall r \in R_d^{\OBL}[q,i]$ either $r$ is discarded for $q$, or \\
  (2) \= $\exists s \in R^\OBL[\non q, j]$ such that\\
      \> (2.1) $s$ is applicable for $\non q$, and \\
      \> (2.2) $\forall t \in R^\OBL[q,k]$, either $t$ is discarded for $q$ or $t\not> s$.
\end{tabbing}
\end{minipage}
\end{definition}
\noindent
Notice that, given the intended correspondence between $\OBL l$ and 
$+\partial_{\OBL} l$ (see Definition~\ref{definition:BodyApplicable}) we will 
refer to ``the derivation of $\OBL l$'' when, strictly speaking, we should use
``the derivation of $+\partial_{\OBL} l$''; similarly for when we say that $\OBL l$ 
has been refuted. 

\begin{example}
Let $D=(F,R,>)$ be a defeasible theory, where $F=\{f_1, f_2, g_2, f_3,\allowbreak f_7\}$, 
$R$ is the set of rules given in Example~\ref{ex:rules}, and ${>}= \set{(r_6,r_2)}$. 
$D$ allows us to draw the following derivation:
\begin{align*}
  (1) & ~{+\partial} f_1 && \text{fact}\\
	  & ~\vdots \\
  (5) & ~{+\partial} f_7 && \text{fact}\\
  (6) & ~{+\partial} d   && \text{from }r_7 \text{ and } R^C[\neg f]=\emptyset\\
  (7) & ~{-\partial_\OBL} a && \text{from } r_4 \text{ applicable and } r_1\not> r_4\\
  (8) & ~{+\partial_\OBL} b && \text{from } r_2 \text{ applicable and } r_7 \text{ discarded}\\
  (9) & ~{+\partial}\neg b && \text{from (7) } r_5 \text{ applicable } {-\partial_\OBL a}\in P(1..8)\\
  (10) & ~{+\partial_\OBL} c && \text{from (8) and (9) } r_2 \text{ applicable for }c \text{ and (7) } r_6 \text{ discarded}
\end{align*}
Steps $P(1)\dots P(5)$: According to clause (1) of Definition~\ref{def:+p} every fact is 
defeasible provable. 

For $P(6)$ we have to satisfy the conditions given in Definition~\ref{def:+p}:
We have that rule $r_7$ is applicable for $d$ (clause 2.2), and $R[\neg d]=\emptyset$ satisfying
clause (2.3) vacuously. 

Step $P(7)$ follows from clauses (2.1) and (2.2) of Definition~\ref{def:-pO}:
Given that we have $+\partial d$ at step $P(6)$, rule $r_4$ is (body)-applicable, and the only 
(prescriptive) rule for $a$, rule $r_1$ is not stonger than $r_4$. 

The conclusion in step $P(8)$ is entailed by Definition~\ref{def:+pO}: rule $r_1$ is 
body-applicable, but not applicable for $b$ at index 2, since we have $a$ at index 1, and 
$-\partial_\OBL a$ at $P(7)$; thus $r_1$ is discarded. However, $r_2$ is applicable for $b$ at 
index 1 (the rule is clearly body-applicable given $A(r_2)\subset F$ and all facts are 
defeasibly provable). Thus, clause (1) holds. For clause (2), $r_7$ is (body)-discarded, 
$\OBL a\in A(r_7)$, we have $-\partial_\OBL a$ at $P(7)$, and there are no other rules 
in $R^\OBL[\neg b]$.

The justification for step $P(9)$ follows from $P(7)$ where we proved $-\partial_\OBL a$; 
thus $r_5$ is (body-)applicable (see item 2 of 
Definition~\ref{definition:BodyApplicable}); in addition  $R^C[b]=\emptyset$. 

Finally, for $P(10)$, as we have already argued $r_2$ is body-applicable, and we can 
use $P(8)$ and $P(9)$ to establish that the rule is applicable for  $c$ at index 2. 
In addition, $P(7)$ allows us to determine that $r_6$ is (body-)discarded since 
$\OBL a\in A(r_6)$, but the step proves $-\partial_\OBL a$ (item 2 of
Definition~\ref{definition:BodyApplicable}, body-discarded part).
\end{example}

We are now ready to provide the proof condition under which a conjunctive 
obligation can be derived. The condition essentially combines two requirements.
First, a conjunction holds only when all the conjuncts hold
(individually). Second, the derivation of one of the
individual obligations does not depend on the violation of the other conjunct.
To achieve this, we determine the line of the proof when the obligation appears. Then we check that the negation of the other elements of the conjunction
does not occur in the previous derivation steps. 
\begin{definition}\label{def:+pand}
The proof condition of \emph{defeasible provability for a conjunctive obligation} 
is

\smallskip \noindent
\begin{minipage}{.95\textwidth}
\begin{tabbing}
If $P(n+1)=+\partial_{\OBL} c_1\wedge \cdots \wedge c_m$, then\\
$\forall c_i$, $1\leq i\leq m$,\\
(1) $+\partial_{\OBL}c_i\in P(1..n)$ and \\
(2) \=if $P(k)= +\partial_{\OBL} c_i$, $k\leq n$, then\+\\ 
	$\forall c_j$, $1\leq j\leq m$ and $c_j\neq c_i$, $+\partial\non c_j\notin P(1..k)$. 
\end{tabbing}
\end{minipage}
\end{definition}
\noindent
Again, the proof condition to refute a conjunctive obligation is obtained by strong 
negation from the condition to derive a conjunctive obligation defeasibly.

\begin{definition}
\label{def:-pand}
The proof condition of \emph{defeasible refutability for a conjunctive obligation} 
is

\noindent
\begin{minipage}{.95\textwidth}
\begin{tabbing}
If $P(n+1)=-\partial_{\OBL}c_1\wedge\cdots\wedge c_m$, then\\
$\exists c_i$, $1\leq i\leq m$, such that either\\
(1) $-\partial_{\OBL} c_i\in P(1..n)$ or\\
(2) \=if $P(k)= +\partial_{\OBL} c_i$, $k\leq n$, then\+\\ 
	$\exists c_j$, $1\leq j\leq m$ such that $c_j\neq c_i$ and $+\partial\non c_j\in P(1..k)$. 
\end{tabbing}
\end{minipage}
\end{definition}
In case of a binary conjunctive obligation, the positive proof condition
boils down to 

\smallskip \noindent
\begin{minipage}{.95\textwidth}
\begin{tabbing}
  $+\partial_{\OBL\wedge}$: If $P(n+1)=+\partial_{\OBL} p\wedge q$ then\\
  (1) $+\partial_{\OBL} p \in P(1..n)$ and\\
  (2) $+\partial_{\OBL} q \in P(1..n)$ and\\
  (3) if $P(k)=+\partial_\OBL p$ $(k\leq n)$, then $+\partial\non q\notin P(1..k)$ and\\
  (4) if $P(k)=+\partial_\OBL q$ $(k\leq n)$, then $+\partial\non p \notin P(1..k)$.
\end{tabbing}
\end{minipage}

\smallskip\noindent
Similarly, for the condition for $-\partial_{\OBL\wedge}$.

Consider a derivation where we have the following steps
\begin{align*}
		 & \vdots\\
	P(x) & ~{+\partial_\OBL a}\\
	P(y) & ~{+\partial \neg a}\\
	P(w) & ~{+\partial_\OBL b}\\
	     & \vdots
\end{align*}
with $x<y<w<n$. Can we add $+\partial_\OBL a\wedge b$ at step $P(n+1)$? 
Condition (1) of Definition~\ref{def:+pand} holds, but condition (2) does not, since 
we have $P(w)=+\partial_\OBL b$ and $P(y)=+\partial\neg a$, with $y<w$. On the contrary, 
if the derivation is
\begin{align*}
		 & \vdots\\
	P(x) & ~{+\partial_\OBL a}\\
	P(y) & ~{+\partial_\OBL b}\\
	P(w) & ~{+\partial\neg a}\\
	     & \vdots
\end{align*}
both conditions hold and we can append $+\partial_\OBL a\wedge b$ to the 
derivation. In the second case, having $+\partial\neg a$ after the step where
we concluded $+\partial_\OBL b$ ensures that the obligation of $b$ does not
depend on the violation of the obligation of $a$. Notice that in the first case, 
the order does not necessarily mean that $\OBL b$ depends on $\neg a$, but 
that the form of the derivation does not allow us to establish the independence
of $\OBL b$ from $\neg a$. 

\medskip
Before proving some theoretical results about the logic, we give some
examples to illustrate its behaviour.

\section{Examples of Pragmatic Oddity Scenarios}
\label{sec:scenarios}

The scenarios in this section display some patterns of instances of Pragmatic 
Oddity and how they are dealt with based on the proof theory defined in the 
previous section. Moreover, as we will see, we use them to show a limitation 
of the proof theory: it introduces some non-determinism given that, in general, 
several derivations are possible and the order of the conclusion in a proof can 
affect what we can prove with specific orders.   

In what follows, we use $\dots\Rightarrow c$ to refer to an 
applicable rule for $c$ where we assume that the elements are not related 
(directly or indirectly) to the other literals used in the examples. 

\paragraph{Compensatory Obligations}
The first case we want to discuss is when the conjunctive obligation 
corresponding to the Pragmatic Oddity has as conjuncts an obligation 
and its compensation. This scenario is illustrated by the rule: 
\[
\dots \Rightarrow_{\OBL} a\otimes b
\]
In this case, when the rule is applicable,  we derive 
$+\partial_\OBL a$. Also, if $+\partial\neg a$ holds 
(signalling that the obligation of $a$ has been violated), 
the rule is applicable for $b$ at index 2 (condition 2 of
Definition~\ref{definition:APPL+pO}), and we can derive 
$+\partial_\OBL b$ (corresponding to $\OBL b$). Thus, 
we have the two individual obligations $\OBL a$ and $\OBL b$, 
but we cannot  derive the conjunctive obligation of $a$ and $b$, 
since the proof condition that allows us to derive $+\partial_\OBL b$ 
explicitly requires that $+\partial\neg a$ has been already 
derived. Accordingly, it is impossible to have the obligation 
of $b$ without the violation of the obligation of $a$. Hence, 
we conclude $-\partial_\OBL a\wedge b$.

\paragraph{Contrary-to-duty}
The second case is when we have a CTD.
The following two rules provide the classical representation of a CTD:
\[
 \dots\Rightarrow_{\OBL}a \qquad\qquad \neg a \Rightarrow_{\OBL}b
\]
In this case, it is possible to have situations when the obligation of $b$
is in force without violating the obligation of $a$, namely, when
$a$ is not obligatory. However, as soon as we have $\OBL a$, we need to derive
$\neg a$ to trigger the derivation of $\OBL b$ 
(Definition~\ref{definition:BodyApplicable}). Similarly to the previous 
case, we have $+\partial_\OBL a$ and $+\partial_\OBL b$, but we cannot 
conclude $+\partial_\OBL a\wedge b$; instead $-\partial_\OBL a\wedge b$ 
holds.

\paragraph{Pragmatic Oddity via Intermediate Concepts}
The situations in the previous two cases can be easily detected by a simple
inspection of the rules involved; nevertheless, there could be more complicated cases. 
Specifically, when the second conjunct does not immediately depend on the 
first conjunct, but it depends on a reasoning chain. 
The following three rules illustrate the simplest 
structure for this case: 
\begin{gather*}
  \dots \Rightarrow_{\OBL} a\\
  \neg a \Rightarrow b\\
  b \Rightarrow_{\OBL} c
\end{gather*}
Here to derive $\OBL c$, we need first to prove $b$. To prove $b$, we require 
that $\neg a$ has already been proved.  Again, it is possible to 
conclude $\OBL a$ and $\OBL c$, but not  $\OBL(a \wedge c)$. 

\paragraph{Negative Support}
In the previous case, the support was through an intermediate concept. 
However, given the non-monotonic nature of Defeasible Deontic Logic, 
we can have cases where the support is not to derive the 
other obligation directly from the violation. The violation prevents the 
derivation of the prohibition (or the permission of the opposite) 
of the other conjunct. Consider the following 
set of rules:\footnote{It is worth noting that, in the theory below, the
rules for $\neg b$ and $\neg c$ can be either defeasible rules or 
defeaters producing the same result as far as the derivation of 
$\OBL(a\wedge b)$ is concerned.}
\begin{gather*}
  \dots \Rightarrow_\OBL a\\
  \dots \Rightarrow_\OBL b\\
  c \Rightarrow_\OBL\neg b\\
  \dots \Rightarrow c\\
  \neg a \defeater \neg c
\end{gather*}
To derive $\OBL b$, we have to ensure that the rule for $\OBL\neg b$
is discarded. This means that $c$ should be rejected (i.e., $-\partial c$). 
We have two options: the rule for $c$ is discarded, or the rule
for $\neg c$ is applicable. This latter implies that to prove $+\partial_\OBL b$ 
we have to prove first $+\partial\neg a$. Thus, one of the two elements 
of the conjunctive obligation $\OBL(a\wedge b)$ depends on the violation
of the other.

\paragraph{Iterated Conjunctive Obligations}

The two previous examples show that the dependency of one of the conjuncts
from the violation can be negative and indirect. Now, the logic allows for
conjunctive obligations in the body of rules, so the intermediate concept 
could be a conjunctive obligation itself (and we have to use the mechanism 
to determine the independence iteratively). Consider the following theory:
\begin{gather*}
  \dots \Rightarrow_\OBL a\\
  r_2\colon \dots \Rightarrow_\OBL b\\
  \OBL(a\wedge b) \Rightarrow_\OBL c\\
  \dots \Rightarrow_\OBL d\\
  \OBL(c\wedge d) \Rightarrow e
\end{gather*} 
Here, to prove $e$, we have to determine if the conjunctive obligation 
$\OBL(c\wedge d)$ holds. Accordingly, we have to show that $\OBL c$ and 
$\OBL d$ are derivable (and neither depends on the violation of the 
other).  For $\OBL c$, the problem reduces to determining whether the 
conjunctive obligation $\OBL(a\wedge b)$ obtains or not, where we have to 
repeat the procedure for $\OBL a$ and $\OBL b$.  Given the theory above, 
there are no dependencies on violations so that we can conclude $e$. Suppose 
that we replace $r_2$ with
\[
r_2'\colon \neg d\Rightarrow_\OBL b
\]
In this situation, we are still able to derive the four individual obligations, 
and the conjunctive obligation $\OBL(a\wedge b)$; however, we are no longer
able to conclude $\OBL(c\wedge d)$ because $\OBL c$ depends (indirectly) 
on the violation of $\OBL d$. 

\paragraph{Multiple Conjuncts}

In the previous scenario, we consider only cases of binary conjunctions.  In this
example, and in the next one, we are going to examine the situation of pragmatic 
oddity with conjunctions involving more than two conjuncts.  The first set of rules
to analyse is: 
\begin{gather*}
  \dots \Rightarrow_\OBL a\\
  \dots \Rightarrow_\OBL b\\
  \neg a, \neg b \Rightarrow_\OBL c
\end{gather*}
Clearly, to derive $\OBL c$ we need both $\neg a$ and $\neg b$; thus, we derive
$-\partial_\OBL a\wedge b\wedge c$, $-\partial_\OBL a \wedge c$ and 
$-\partial_\OBL b\wedge c$. Finally, as far as conjunctive obligations are concerned
we can conclude $+\partial_\OBL a\wedge b$, noticing that $\OBL(a\wedge b)$ is not 
a Pragmatic Oddity instance.

\paragraph{Multiple Dependencies}
In contrast to the example we just examined where $\OBL c$ depended on the 
conjunction of the two violations, what if it depends on them disjunctively?
Thus, we have the following theory. 
\begin{gather*}
  r_1\colon\dots \Rightarrow_\OBL a\\
  r_2\colon\dots \Rightarrow_\OBL b\\
  r_3\colon\neg a\Rightarrow_\OBL c\\
  r_4\colon\neg b\Rightarrow_\OBL c
\end{gather*}
Let us consider the derivation below:
\begin{align*}
  (1) & ~{+\partial} \neg a && \text{fact}\\
  (2) & ~{+\partial_\OBL} a && \text{from }r_1\\
  (3) & ~{+\partial_\OBL} b && \text{from }r_2\\
  (4) & ~{+\partial_\OBL} c && \text{from (1) and }r_3\\
  (5) & ~{-\partial_\OBL} a \wedge b \wedge c && \text{from (1)--(4), } \neg a\in P(1..4)\\
  (6) & ~{+\partial} \neg b && \text{fact}\\
  (7) & ~{+\partial_\OBL} b\wedge c && \text{from (3) and (4), } \neg b\notin P(1..4)\\
  (8) & ~{-\partial_\OBL} a\wedge c && \text{from (1) and (4), } \neg a\in P(1..4)
\end{align*}
We can carry out a similar proof by swapping the positions of $\neg a$ and $\neg b$,
using $r_4$ in step (4) --yielding $-\partial_\OBL a\wedge b\wedge c$, and 
$-\partial_\OBL b\wedge c$--, but proving $\OBL(a\wedge c)$. Hence, we have a situation
where it is impossible to prove $\OBL(a\wedge b\wedge c)$, but we can prove both
$\OBL(a\wedge b)$ and $\OBL(b\wedge c)$, though it is impossible to have both of them 
in a single derivation.    

\paragraph{Pragmatic Un-pragmatic Oddity}

What about when there are multiple norms both prescribing the 
contrary-to-duty obligation and at least one of the norms is not related 
to the violation of the primary norm?
\begin{gather*}
  r_1\colon \dots \Rightarrow_{\OBL} a \otimes b\\
  r_2\colon \dots \Rightarrow_{\OBL} b\\
  \neg a
\end{gather*}
In this situation you can have a derivation:
\begin{align*}
  &(1)~{+\partial}\neg a && \text{fact}\\
  &(2)~{+\partial_{\OBL}} a && \text{from }r_1\\
  &(2)~{+\partial_{\OBL}} b && \text{from }r_1\text{ and (1) and (2)}
\end{align*}
where the derivation of $\OBL b$ ($+\partial_\OBL b$) depends on the
violation of the primary obligation of $r_1$. In this case, we cannot derive
the conjunctive obligation of $a$ and $b$.  However, there is an 
alternative derivation, namely:
\begin{align*}
  &(1)~{+\partial_{\OBL}} a  && \text{from }r_1\\
  &(2)~{+\partial_{\OBL}} b  && \text{from }r_2\\
  &(3)~{+\partial\neg} a     && \text{fact}\\
  &(4)~{+\partial_{\OBL}} a\wedge b && \text{from (1) and (2)}
\end{align*}
The proof demonstrates the independence of $\OBL b$ from $\neg a$, given 
that the derivation of $\neg a$ occurs in a line after the line
where $+\partial_{\OBL}b$ is derived. 

\paragraph{Iterated Un-pragmatic Pragmatic Oddity}
We have seen cases where multiple derivations
are possible, leading to opposite results about the derivability of instances 
of conjunctive obligations (irrespective of whether they are pragmatic oddity instances).  
Furthermore, a conjunctive obligation can depend on a pragmatic oddity instance. 
For example, the following set of rules illustrates a situation where we have a derivation 
refuting an instance of pragmatic oddity, and a second one where the same instance 
is derivable. In turn, this instance can make a conjunctive obligation derivable
or not.   
\begin{gather*}
  r_1\colon\dots \Rightarrow_\OBL a\otimes b\\
  r_2\colon\dots \Rightarrow_\OBL a\\
  r_3\colon\dots\Rightarrow_\OBL b\\
  r_4\colon\dots \Rightarrow \neg a\\
  r_5\colon \OBL(a\wedge b) \Rightarrow_\OBL \neg c\\
  r_6\colon\dots \Rightarrow_\OBL c\otimes d\\
  r_7\colon\dots \Rightarrow_\OBL d
\end{gather*}
\allowdisplaybreaks
The key point of this example is that we have a conjunctive obligation, 
$\OBL(a \wedge b)$, in the antecedent of a prescriptive rule, $r_5$, and there 
is a second rule, $r_6$, for the opposite of the conclusion of $r_5$.  
\begin{align*}
  (1) & ~{+\partial} \neg a &&  \text{from } r_4\\
  (2) & ~{+\partial_\OBL} a &&  \text{from } r_2 \text{ or } r_1\\
  (3) & ~{+\partial_\OBL} b &&  \text{from } r_1 \text{ or } r_3\\
  (4) & ~{-\partial_\OBL} a\wedge b && \text{from (1), }{+}\partial\neg a\in P(1..3)\\
  (5) & ~{-\partial_\OBL} \neg c && \text{from (4) and } r_5\\
  (6) & ~{+\partial_\OBL} c && \text{from } r_6\\ 
  (7) & ~{+\partial_\OBL} d && \text{from } r_7\\
  (8) & ~{+\partial_\OBL} c\wedge d && \text{from (6) and (7), } {+\partial}\neg c, +\partial\neg d\notin P(1..7)
\end{align*}
This proof blocks the derivation of $\OBL(a\wedge b)$, since $+\partial\neg a$ occurs in $P$ before $+\partial_\OBL b$. Consequently, we can derive the conjunctive obligation
$\OBL(c \wedge d)$, since rule $r_5$ is discarded. However, if we postpone
the use of $r_4$, namely, doing the proof with the sequence
\begin{align*}
  (1) & ~{+\partial_\OBL} a && \text{from } r_1 \text{ or } r_2\\
  (2) & ~{+\partial_\OBL} b && \text{from } r_3\\
  (3) & ~{+\partial_\OBL} a\wedge b && \text{from (1) and (2), } {+}\partial\neg a,+\partial\neg b\notin P(1..2)\\
  (4) & ~{-\partial_\OBL} c &&\text{from (3) and } r_5\\
  (5) & ~{-\partial_\OBL} c\wedge d && \text{from (4)}
\end{align*}
we are allowed to derive $+\partial_\OBL a\wedge b$, making $r_5$ applicable,
preventing the derivation of $\OBL c$. 
Suppose that $r_5$, instead of being a prescriptive rule, is a constitutive rule,
namely
\[
r_5\colon \OBL(a\wedge b) \Rightarrow \neg c
\]
enabling us to prove or refute the violation of the first element of $r_6$. 
Using the first derivation in step (5) we conclude $+\partial\neg c$. Now, we
have two ways to derive $\OBL d$: using $r_6$ (leading to an instance of pragmatic 
oddity), or using $r_7$, and we can postpone the derivation of $\neg c$, 
allowing us to assert $\OBL(c \wedge d)$. 

\paragraph{Mix and Match}
In all the previous cases the focus was on conjunctions where 
one of the conjuncts somehow depended on one of the other 
conjuncts. In other terms, the conjunction contains a primary
obligation and a secondary obligation (an obligation in 
force after the violation of another obligation). Consider the 
rules
\begin{gather*}
	r_1\colon \dots \To_\OBL a \otimes b\\
	r_2\colon \dots \To_\OBL c\otimes d\\
	r_3\colon \dots\To \neg a\\
	r_4\colon \dots\To \neg c
\end{gather*}
Here, from $r_1$ we obtain $\OBL a$ ($+\partial_\OBL a$); similarly
from $r_2$ we get $\OBL c$. $r_3$ and $r_4$ allow us to derive the 
literals corresponding to the violations of the two obligations,
namely $+\partial\neg a$ and $+\partial \neg c$. Now, $r_1$ and 
$r_2$ applicable for their element at index 2. Hence, we conclude
$\OBL b$ ($+\partial_\OBL b$) and $\OBL d$ ($+\partial_\OBL d$); can we 
derive their conjunctive obligation? The answer is positive. 
$\OBL b$ does not depend on the violation of $\OBL d$ (there is no way 
to derive $\neg d$ from the rules above) and the other way around. 
$\OBL(b \wedge d)$ is a conjunctive obligation of two secondary 
obligations, what about other conjunctions, e.g., $\OBL(a\wedge d)$ and 
$\OBL(c\wedge b)$? Again the answer is positive: there is 
no need to derive $+\partial\neg a$ for the derivation 
of $+\partial_\OBL d$; the argument for the second is the same.

\section{A Bottom-up Characterisation}
\label{sec:bottom-up}

The examples in the previous section illustrate cases 
where multiple derivations are possible and whether a conjunctive 
obligation is derivable or not depends on the specific derivation. 
Furthermore, the non-monotonicity of the logic presents other complications.
Whether some conclusions are derivable depends on other 
elements being derivable, and these depend on specific derivations. 
Hence, we need to devise a mechanism that does not rely on a particular 
order in which a derivation sequence is laid out. The idea of
the proof conditions for conjunctive obligations is to see that in 
the derivation of an obligation, the derivation of the violation of
the other conjunct does not appear. Alternatively, we can say that
an obligation is independent of the violation if we can push 
down in the proof the derivation of the violation. If the derivation 
is independent, then the rules to derive the violation do not 
contribute to the derivation of the obligation. Consequently, we could
remove such rules without affecting the derivability of the obligation. 
Given that the derivations of the obligations of the conjuncts in a 
conjunctive obligation must be independent of the derivation of violations 
of those obligations and that when they are independent, we can run 
the derivations in parallel (using separate subsets of the rules), 
then we can inquire whether it is possible to carry out these derivations 
in a single construction. The answer is positive, 
and we can adapt the bottom-up construction of Maher and 
Governatori~\cite{Maher:1999:A-Semantic}.  The idea of the bottom-up
construction is that instead of working in a goal-directed fashion, 
we work in stages. For each stage, we determine all the conclusions
that can be ``derived'' at that stage, assuming that whatever was
in a previous stage is already provable. Accordingly, we start from 
the empty set, and in the first stage, we determine what is provable
from the empty set; then, for stage $n+1$, we see what is provable from 
stage $n$.  Before defining the extension of
a defeasible theory, we need to provide a mechanism that 
guarantees that the derivation of an obligation does not depend
on the violation of another obligation. To this end, we introduce a 
construction, called reduction, that removes all rules 
for a particular element from a theory. In what follows, we are going to use 
the reduct to remove all rules for the violation of an obligation,
and we are going to examine whether the other obligation is still derivable
from the reduced theory. If it does, then the obligation does not 
depend on the violation. 

\begin{definition}
\label{def:reduct}
Given a defeasible theory $D=(\FACTS,R,>)$ and a set of (plain) literals 
$L=\set{l_1,l_2,\dots}$, the 
\emph{reduct of $D$ based on $L$}, 
noted as $\reduct(D,L)$ is the defeasible theory $D'=(\FACTS',R',>)$ 
satisfying the following conditions:
\begin{enumerate}
  \item $\FACTS'=\FACTS\setminus L$;
  \item $R'=R\setminus \bigcup_{l\in L}R[l]$;
  \item ${>'}={>} \setminus\set{(r,s)\colon r\notin R' \vee s\notin R'}.$
\end{enumerate}  
\end{definition}
The idea of the transformation is to create a theory similar to the 
original theory, as we said, without the literals in $L$. The condition on $F$ is obvious. 
The second condition ensures that for each literal $l\in L$ the rules that can derive the literal 
are removed. Then the literal is no longer derivable since the 
resulting theory does not contain rules for the literal anymore. Given
that $R'[l]=\emptyset$, the following result is immediate.

\begin{observation}
Given a Defeasible Theory $D$, and a set $L$ of literals, $-\partial l$ is 
derivable in $\reduct(D,L)$ for $l\in L$.
\end{observation}
It is worth noting that we do not have to remove rules where the literals in $L$
appear in the antecedent of the rule. Such rules are immediately 
discarded.  Similarly, for prescriptive rules where the complement 
of the removed literals appears in the head of the rules. Such rules 
are no longer applicable for elements appearing after one of the removed literals. 
Thus if you have a rule with the $\otimes$-chain 
$c_1\otimes\cdots\otimes c_n\otimes \neg l\otimes c_{n+1}\cdots$, the rule is
in $R^{\OBL}[c,m]$, but it is not applicable for any $m\geq n+1$. Remember that 
to derive $+\partial_{\OBL}c_{n+1}$ we have to prove both $+\partial_{\OBL}\neg l$ 
and $+\partial l$.

\begin{example}
Consider a theory $D$ whose set of rules $R$ consists of the rules presented 
in the Iterated Un-pragmatic Pragmatic Oddity scenario described in the 
previous section including the additional rule 
$r'_5\colon \OBL(a \wedge b) \Rightarrow \neg c$.  The reduct of $D$ based on 
$L=\set{\neg a}$, $\reduct(D,\set{\neg a})$ has the following set of rules 
$\set{r_1, r_2, r_3, r_5, r'_5, r_6, r_7}$.  For $\reduct(D,\set{\neg g})$, 
$R'=\{r_1, r_2, r_3, r_4, r_5, r_6, r_7\}$.  Finally, for the reduct based on 
$L=\set{\neg a, \neg c}$, the resulting set of rules is $\{r_1, r_2, r_3, r_5,$
$r_6, r_7\}$.  Given that we removed rules for the elements in $L$, those literals cannot be derived positively; indeed, we derive 
$-\partial\neg a$ in $\reduct(D,\set{\neg a})$, $-\partial\neg c$ in 
$\reduct(D,\set{\neg c})$, and both of them in $\reduct(D,\set{\neg a, \neg c})$; 
\end{example}

We can now specify when a (deontic) literal is independent of a set of  
plain literals in Defeasible Deontic Logic
\begin{definition}\label{def:independence}
  Given a defeasible theory $D$, a set $L$ of plain literals, and a literal 
  $m$, $m$ is \emph{independent from} $L$ iff $m$ is defeasibly provable in $D$ 
  and in $\reduct(D,L)$. 
\end{definition}
We can now show that condition (2) in the proof conditions for 
a conjunctive obligation ensures the independence of the obligations
from the violations.  However, before proving this result, we have 
to recall a general property about Defeasible (Deontic) Logic:
First of all, a defeasible theory is consistent if $\FACTS$ does not
contain a literal $l$ and its complement $\neg l$.  Second, given 
a logical formula expressing a proof condition of the strong negation 
of the formula/conditions is obtained by replacing every occurrence of 
a positive proof tag with the corresponding negative proof tag, 
replacing conjunctions with disjunctions, disjunctions with conjunctions,
existential with universal and universal with existential.    
Is it immediate to observe that each negative proof condition given
in Section~\ref{sec:deflogic} is the strong negation of the corresponding 
positive one (and the other way around).  

If corresponding proof conditions are defined using the principle of 
strong negation outlined above, then, given a derivation, it is 
impossible to have that the literal (conjunctive obligation) is both 
derivable and refutable in the same derivation. 
\begin{proposition}\label{prop:consistency}\cite{igpl09policy}
  Given a consistent defeasible theory $D$, a derivation $P$, a literal $l$, 
  and  proof tag $\#\in\set{\partial,\partial_{\OBL}}$ it is
  not possible that $+\# l,-\#l\in P$. 
\end{proposition}
\begin{proof}
The proof is an extension of the proof given in \cite{igpl09policy}. 
\cite{igpl09policy} proves that if the proof conditions for a pair of 
proof tags $+\#$ and $-\#$ are defined as the strong negation of each other,
there is no theory $D$ such that $+\# l$ and $-\# l$ both hold. The proof 
conditions for literals are the same as those in \cite{jpl:permission} and the 
result applies to them; and if the property holds for theories, it holds 
for individual proofs as well. The proof conditions for conjunctive 
obligations extend the constraints in \cite{igpl09policy} 
since they require that some elements are not in a derivation (clause (2)). 
Let us consider the case of the proof conditions for conjunctive 
obligations (Definitions~\ref{def:+pand} and \ref{def:-pand}). 
Suppose we have a derivation where we have both $+\partial_\OBL \seq{c}$ 
at step $n$ and $-\partial_\OBL \seq{c}$ at step $m$. Let us assume 
that $m<n$ (the case $n<m$ is analogous). Thus, by clause (1) of Definition~\ref{def:+pand} we have
that for all $c_i$, $+\partial_\OBL c_i\in P(1..n)$ (if an element is 
in $P(1..m)$ and $m<n$, the element is also in $P(1..n)$); if clause (1) of 
Definition~\ref{def:-pand} holds, then there is a $c_i$ such that
$-\partial_\OBL c_i\in P(1..n)$, contradicting the results for the other 
proof conditions. Thus clause (2) of Definition~\ref{def:+pand} must hold. Let $k$ be the step 
where we derived $+\partial_\OBL c_i$. Now, $k\leq m<n$, and we have 
$+\partial\non c_j \in P(1..k)$, while for clause (2) of Definition~\ref{def:+pand}, 
it should be $+\partial\non c_j\notin P(1..k)$, thus even in this case
we obtain a contradiction. Accordingly, in any case we have a contradiction; thus,
it is impossible to derive and refute a conjunctive obligation in the 
same derivation.
\end{proof}

Armed with this result, we can prove the result linking independence
and the proof conditions for conjunctive obligations. 
\begin{proposition}\label{prop:independence}
  Given a consistent defeasible theory $D$, a deontic literal 
  $m$ and a set $L$ of plain literals. $m$ is independent from $L$ iff 
  there is a derivation $P$ in $D$ such that
  \begin{enumerate}
    \item $P(n)=+\partial_{\OBL}m$ and
    \item $\forall l\in L$, $+\partial l\notin P(1..n)$.
  \end{enumerate}   
\end{proposition}
\begin{proof}
By the definition of independence (Definition~\ref{def:independence}) there
is a derivation $P'$ in $\reduct(D,L)$ for $+\partial_\OBL m$. 
By construction of $\reduct(D,L)$, for $l\in L$ $R[l]=\emptyset$; thus,
we can add $-\partial l$ to any derivation in $\reduct(D,L)$; by 
Proposition~\ref{prop:consistency} $+\partial l$ is not derivable, and 
no derivation in $\reduct(D,L)$ can contain it.  What we have to do,
is to show how to ensure that the derivation in $\reduct(D,L)$ guarantees
that there is a derivation in $D$. More specifically, we are going to give 
a (constructive) procedure to transform $P'$ into a proof in $D$. The 
procedure removes all steps that do not contribute to the derivation of $m$.
Let us assume that $P'(k)=+\partial_\OBL m$. We now consider the conclusion 
in $P'(k-1)$. If it is a justification for step $P'(k)$, we keep it; 
otherwise, we delete it.  A conclusion (tagged literal or tagged conjunction) 
is a justification for a step of a derivation $P(n)$ if it makes discarded 
a rule attacking the conclusion in  $P(n)$ (see Definition~\ref{definition:DISC+pO+pP}), 
or it contributes to making applicable  a rule for the conclusion in $P(n)$ 
(see Definition~\ref{definition:APPL+pO}).  We repeat step by step backward
the procedure for all step $P'(k-z)$ to determine if they justify the step $P'(w)$, $k-z<w\leq k$ that have not been deleted
in the previous iterations of the procedure. It is clear that the resulting sequence 
is still a proof for $+\partial_\OBL m$ in $\reduct(D,L)$, and thus does not 
contain any step of the form $+\partial l$ for $l\in L$. Finally, since the 
rules used to check if the remaining steps are a subset of the rules in $D$, 
and again, if it does not involve any step with $+\partial l$, then the proof is 
also a proof for $+\partial_\OBL m$ in $D$.

For the other direction, a proof in $D$ for $+\partial_\OBL m$ that 
does not contain any step of the form $+\partial l$ is trivially a proof in $D$.
The rules missing in $\reduct(D,L)$ are the rules in $R[l]$; these rules would
be used to justify steps of the form $+\partial l$ (which are not present in 
$P$ anyway); thus, the proof $P$ is also a proof in $\reduct(D,L)$.
\end{proof}

\begin{example}
Consider again a theory $D$ containing the rules for the Iterated Unpragmatic Pragmatic Oddity 
scenario. Notice that the second derivation provided in that section is 
effectively a derivation in $\reduct(D,\set{\neg a})$, and we can use it 
to show that we can derive $+\partial_\OBL b$ in $\reduct(D,\set{\neg a})$. 
At the same time, the derivation is also a derivation in $D$, and thus it 
shows the independence of $\OBL a$ from $\neg a$. Hence, we can conclude 
$\OBL(a\wedge b)$. 

Suppose we replace $r_5$ with its constitutive version, i.e., 
$r'_5\colon \OBL(A\wedge b)\To \neg c$. Since we no longer have a 
prescriptive rule for $\OBL\neg c$, rule $r_6$ is unopposed and we can 
derive $+\partial_\OBL c$ from it. Also, $+\partial\neg c$ and $+\partial_\OBL d$ 
are derivable, and we can ask if the instance of the pragmatic oddity $\OBL(c \wedge d)$ 
is derivable.   To this end, we compute the reduct $\reduct(D,\set{\neg c})$. 
In this theory we do not have $r'_5$, and it is easy to check that we can 
derive both $\OBL c$ and $\OBL d$. Accordingly, $\OBL d$ does not depend 
on $\neg c$, and we can conclude $+\partial_\OBL c \wedge d$.

Notice that we can combine the two independence results, and we are still able 
to conclude $+\partial_\OBL d$ when we remove the rules for $\neg a$ and $\neg c$. 
Finally, strictly speaking, to prove the two conjunctive obligations $\OBL(a\wedge b)$ 
and $\OBL(c\wedge d)$ we have to show that $\OBL a$ and $\OBL c$ are independent of, 
respectively, $\neg b$ and $\neg d$. However, since there are no constitutive rules 
for $\neg b$ and $\neg d$, the theory $D$ itself is its reduct, that is 
$\reduct(D,\set{\neg b})=\reduct(D,\set{\neg d})=\reduct(D,\set{\neg b,\neg d}) = D$.
\end{example}

We are now ready to introduce the notion of extension. Typically in 
Defeasible Logic, the extension of a theory is the set of all the literals
that can be derived from the theory. However, in Defeasible Deontic Logic,
the order of the elements in a derivation does not matter. On the contrary,
as we have seen, the order matters if we want to capture the pragmatic 
oddity phenomenon properly, and different derivations are possible. In the definition below,
we continue to speak of derivable/refutable literals/conjunctions. 
The fixed point construction will give the precise notion of derivation/refutation in Definition~\ref{def:fix-point}. Formally, the extension is a 
6-tuple of sets, where every set contains the derivable/rejected 
literals/conjunctive obligations.

\begin{definition}\label{def:extension}
  Given a defeasible theory $D$ the \emph{extension} ($E(D)$) of the 
  theory is the tuple: 
 \[
 E(D)=
\langle 
  \partial^+(D), \partial^-(D),\partial^{+\OBL}(D),\partial^{-\OBL}(D),
  \partial^{+\wedge}(D), \partial^{-\wedge}(D)
\rangle
 \] 
where
\begin{itemize}
  \item $\partial^+(D)$ is the set of literals appearing in $D$ that 
    are defeasible provable as institutional statements;
  \item $\partial^-(D)$ is the set of literals appearing in $D$ that 
    are defeasible refutable as institutional statements;
  \item $\partial^{+\OBL}(D)$ is the set of literals appearing in $D$ that 
    are defeasible provable as obligations;
  \item $\partial^{-\OBL}(D)$ is the set of literals appearing in $D$ that 
    are defeasible refutable as obligations;
  \item $\partial^{+\wedge}(D)$ is the set of conjunctive obligations 
  	defeasibly provable in $D$ whose conjuncts are literals appearing in $D$; 
  \item $\partial^{-\wedge}(D)$ is the set of conjunctive obligations 
 	defeasibly refutable in $D$ whose conjuncts are literals appearing in $D$.
\end{itemize}
\end{definition}
Before we move to the procedure to construct the extension of a given 
defeasible theory, we need some auxiliary definitions (these definitions are 
the counterpart of Definitions 
\ref{definition:BodyApplicable}--\ref{definition:DISC+pO+pP} 
for extensions instead of derivations).  First of all, we 
introduce some notation to identify the types of literal occurring
in the body of rules. 

\begin{definition}\label{def:set-in-rules}
Given a rule $r$, we identify the following sets of literals (and conjunctions 
of literals).
\begin{itemize}
  \item $r^\#=\set{l\in\LIT\colon l\in A(r)}$;
  \item $r^\OBL=\set{l\in\LIT\colon \OBL l\in A(r)}$;
  \item $r^\PERM=\set{l\in\LIT\colon \neg\OBL l\in A(r)}$;
  \item $r^\wedge=\set{c=l_1\wedge\cdots\wedge l_n\colon \OBL(c)\in A(r)}$.
\end{itemize}
\end{definition}
\begin{example}
Consider the rule 
\begin{equation}\label{eq:formuala}
r\colon	a, \neg b, \OBL\neg c, \neg\OBL a, \OBL(c\wedge \neg d) \To_\OBL e \otimes f
\end{equation}
Here, $r^\#=\set{a,\neg b}$, $r^\OBL=\set{\neg c}$, $r^\PERM=\set{a}$ and $r^\wedge=\set{c\wedge \neg d}$.
\end{example}
Given a conjunction $c_1\wedge\dots\wedge c_m$,  we use $C$ to denote the set of 
complements of the literals in the conjunction, namely: $C=\set{\non c_1,\dots,\non c_m}$.

The next three definitions just mimic the definitions of when 
rules are applicable or rejected; instead of applying to steps 
of a derivation, they apply to elements of an extension.
In the construction we are going to use to compute the extension of 
a theory, the reference is to the previous stage of the construction
of the extension.
\begin{definition}\label{definition:BodyApplicable-e}
	A rule $r \in R[q,j]$ is \emph{body-applicable in an extension $E(D)$} iff 
\begin{enumerate}
  \item $r^\#\subseteq\partial^+(D)$ and
  \item $r^\OBL\subseteq\partial^{+\OBL}(D)$ and
  \item $r^\PERM\subseteq\partial^{-\OBL}(D)$ and
  \item $r^\wedge\subseteq\partial^{+\wedge}(D)$.
\end{enumerate}
A rule $r \in R[q,j]$ is \emph{body-discarded in an extension $E(D)$} iff 
\begin{enumerate}
  \item $r^\# \cap \partial^{-}(D)\neq\emptyset$ or
  \item $r^\OBL \cap \partial^{-\OBL}(D)\neq\emptyset$ or
  \item $r^\PERM \cap \partial^{+\OBL}(D)\neq\emptyset$ or
  \item $r^\wedge \cap \partial^{-\wedge}(D)\neq\emptyset$.
\end{enumerate}
\end{definition}

\begin{definition}\label{definition:APPL+pO-e}
A rule $r \in R^\OBL[q,j]$ such that $C(r) = c_{1} \otimes \cdots \otimes c_{n}$ is \emph{applicable in an extension $E(D)$} for literal $q$ at
index $j$, with $1 \leq j < n$, in the condition for $\partial^{\pm\OBL}$ iff 
\begin{enumerate}[nosep]
  \item $r$ is body-applicable in $E(D)$; and
  \item for all $c_k \in C(r)$, $1 \leq k < j$, $c_k\in\partial^{+\OBL}(D)$ and $\non c_k \in\partial^{+}(D)$.
\end{enumerate}
\end{definition}

\begin{definition}\label{definition:DISC+pO+pP-e}
A rule $r \in R[q,j]$ such that $C(r) = c_{1} \otimes \cdots \otimes c_{n}$ is \emph{discarded in an extension $E(D)$} for literal $q$ at
index $j$, with $1 \leq j \leq n$ in the condition for $\partial^{\pm\OBL}$ iff
\begin{enumerate}[nosep]
  \item $r$ is body-discarded in $E(D)$; or 
  \item there exists $c_k \in C(r)$, $1 \leq k < l$, such that either $c_k\in\partial^{-\OBL}(D)$ or $\non c_k\in\partial^-(D)$. 
\end{enumerate}
\end{definition}

According to Definition~\ref{definition:BodyApplicable} a rule $r$ is (body-)applicable 
if all the elements in the antecedent of the rule $A(r)$ have been proved 
in previous steps of the derivation. Similarly, $r$ is (body-)discarded if there 
is an element in the antecedent that has been refuted.  The idea behind the construction 
of the extension of a theory is to start from the set of facts and derive all 
conclusions (positive and negative) that can be obtained directly from the facts. 
Then, the procedure works as follows: At every iteration, we compute all the conclusions
that follow directly from the elements calculated in the previous extension. 
A key aspect is determining what rules are applicable or discarded at a particular 
iteration. 
\begin{example}
When we consider again the rule $r$ in \eqref{eq:formuala}, 
then $r$ is applicable in an extension 
$E(D)$, if $\set{a, \neg b}\subseteq \partial^+(D)$, $\set{\neg c}\subseteq\partial^\OBL(D)$,
$\set{a}\subseteq\partial^\PERM(D)$ and $\set{c\wedge \neg d}\subseteq\partial^\wedge(D)$.
In addition, to check if it is applicable for $f$ at index 2, we have to see if $e\in\partial^\OBL(D)$ and $\neg e\in\partial^-(D)$. 
The rule is discarded if one of the given sets has a non-empty intersection with the corresponding 
negative sub-part of the extension, indicating, in this case, that one of the elements 
has been refuted. 
\end{example}

We are now ready to give the definition providing the procedure
to compute the extension of a defeasible theory. 
\begin{definition}[Extension Construction]\label{def:fix-point}
Given a defeasible theory $D=(F,R,>)$ the extension of $D$ is built by the 
following construction
\begin{align*}
E_{n+1}(D)=&\langle
  \partial^+_{n+1}(D),
  \partial^-_{n+1}(D),
  \partial^{+\OBL}_{n+1}(D),
  \partial^{-\OBL}_{n+1}(D),
  \partial^{+\wedge}_{n+1}(D),
  \partial^{-\wedge}_{n+1}(D)
\rangle\\
=&
\langle
  \mathcal{T}(\partial^+_{n}(D)),
  \mathcal{T}(\partial^-_{n}(D)),
  \mathcal{T}(\partial^{+\OBL}_{n}(D)),\\
  &\phantom{\langle}
  \mathcal{T}(\partial^{-\OBL}_{n}(D)),
  \mathcal{T}(\partial^{+\wedge}_{n}(D)),
  \mathcal{T}(\partial^{-\wedge}_{n}(D))
\rangle
\end{align*}
where
\[
 E_0(D)=\langle F,\emptyset,\emptyset,\emptyset,\emptyset,\emptyset\rangle
\]
and
\begin{tabbing}
  $\mathcal{T}(\partial^+_{n}(D))$\= ${}= \partial^{+}_n \cup
    \{q\colon$\= $\non q\notin F$ and\+\+\\ 
         $\exists r\in R_d[q]$ $r$ is applicable in $E_n(D)$ and \\
         \phantom{xx}$\forall s\in R[\non q]$ $s$ is either discarded in $E_n(D)$ or\\
         \phantom{xxxx}$\exists t\in R[q]$ $t$ is applicable in $E_n(D)$ and $t>s\}$\-\-\\ 
  $\mathcal{T}(\partial^-_{n}(D)) = \partial^{-}_n \cup 
    \{q\colon \non q\in F$ or\+\+\\ 
    $\forall r\in R_d[q]$ either $r$ is discarded in $E_n(D)$ or\\
    \phantom{xx}$\exists s\in R[\non q]$ $s$ is applicable in $E_n(D)$ and\\
    \phantom{xxxx}$\forall t\in R[q]$  either $t$ is discarded in $E_n(D)$ or $t\not>s\}$\-\-\\
  $\mathcal{T}(\partial^{+\OBL}_{n}(D)) = \partial^{+\OBL}_n \cup
      \{q\colon$\= $\exists r\in R^\OBL[q,j]$ $r$ is applicable in $E_n(D)$ and\+\+\\
      \phantom{xxxx}$\forall s\in R^\OBL[\non q,k]$ $s$ is either discarded in $E_n(D)$ or\\
      \phantom{xxxxxx}$\exists t\in R^\OBL[q,m]$ $t$ is applicable in $E_n(D)$ and $t>s\}$\-\-\\ 
  $\mathcal{T}(\partial^{-\OBL}_{n}(D)) = \partial^{-\OBL}_n \cup
    \{q\colon \forall r\in R^\OBL[q,j]$ either $r$ is discarded in $E_n(D)$ or\+\+\\
       \phantom{xxxx}$\exists s\in R^\OBL[\non q,k]$ $s$ is applicable in $E_n(D)$ and\\
     \phantom{xxxxxx}$\forall t\in R^\OBL[q,m]$  either $t$ is discarded in $E_n(D)$ or $t\not>s\}$\-\-\\
  $\mathcal{T}(\partial^{+\wedge}_{n}(D)) = \partial^{+\wedge}_n \cup 
  \{c_1\wedge\dots\wedge c_m\colon \forall c_i$, 
      \=$c_i\in \partial_n^{+\OBL}$ and \+\\
        $c_i \in \partial^{+\OBL}(\reduct(D,C\setminus\set{\non c_i}))\}$\-\\
$\mathcal{T}(\partial^{-\wedge}_{n}(D)) =\partial^{-\wedge}_n \cup
  \{c_1\wedge\dots\wedge c_m\colon \exists c_i$, $c_i\in\partial_n^{-\OBL}(D)$ or \+\\
  $c_i \notin \partial^{+\OBL}(red(D,C\setminus \set{\non c_i}))\}$
\end{tabbing}
\end{definition}
In the construction above, the first four sets replicate the proof 
conditions for the corresponding proof tags where we proceed in terms of 
stages instead of derivation steps. For conjunctive obligations, 
we first determine if the individual obligations are derivable at the 
current stage. At the same time, for each individual obligation,
we check if it is in the extension of the reduct of the theory obtained by 
removing the literals corresponding to the violations of the other obligations 
in the conjunctive obligation. If it is, then 
the individual obligation is independent of the violation of the other 
obligations. Notice that for this last step, we are not looking if 
they are in the extension in a particular stage but in the extension 
at the end of the construction for the extension of the reduct. Also, 
further reducts (for
other conjunctions) may be computed in the computation for a reduct.  However, given that a reduct is a subset
of a given theory, the process is guaranteed to terminate (provided that the initial theory has finitely many rules). 
 
The set of extensions forms a complete lattice under the pointwise containment
ordering, with $E_0$ as its least element. The least upper bound operation is
the pointwise union. It is easy to see that $\mathcal{T}$ is monotonic, and
the Kleene sequence from $E_0$ is increasing. Thus the limit 
\[
  \mathcal{L} = \langle
\partial^+_{L}(D),
\partial^-_{L}(D),
\partial^{+\OBL}_{L}(D),
\partial^{-\OBL}_{L}(D),
\partial^{+\wedge}_{L}(D),
\partial^{-\wedge}_{L}(D)
\rangle.
\] 
of all finite elements in the sequence exists, 
and it has a least fixpoint 
\[
  \mathcal{F} = \langle
\partial^+_{F}(D),
\partial^-_{F}(D),
\partial^{+\OBL}_{F}(D),
\partial^{-\OBL}_{F}(D),
\partial^{+\wedge}_{F}(D),
\partial^{-\wedge}_{F}(D)
\rangle
\] 
When $D$ is a finite propositional defeasible deontic theory $\mathcal{F} = \mathcal{L}$.
Accordingly, we take $\mathcal{F}$ as the extension of $D$, $E(D)=\mathcal{F}$. Furthermore, $\mathcal{F}$ being 
the least upper bound is unique and captures the conditions that determine whether a conjunctive 
obligation is independent of the violations of its conjuncts.

We can revisit some of the scenarios presented in Section~\ref{sec:scenarios} 
using the bottom-up construction.  

\begin{example}\label{ex:bu-example}
Let us consider again the theory $D=(F,R,\emptyset)$ we 
used to illustrate the Iterated Pragmatic Non Pragmatic Oddity scenario, 
where $F=\{f_1, f_2, f_3,\allowbreak f_4,\allowbreak f_6, f_7\}$, and $R$ consists of 
the following rules:
\begin{gather*}
	r_1\colon f_1 \To_\OBL a \otimes b \qquad
	r_2\colon f_2 \To_\OBL a \qquad
	r_3\colon f_3 \To_\OBL b \qquad
	r_4\colon f_4 \To \neg a\\
	r_5\colon \OBL(a \wedge b)\To_\OBL \neg c \qquad
	r_6\colon f_6 \To_\OBL c\otimes d \qquad
	r_7\colon f_7 \To_\OBL d.
\end{gather*}
According to Definition~\ref{def:extension}
\[
 E_0(D) = (\set{f_1, f_2, f_3, f_4, f_6, f_7}, 
 		\emptyset, \emptyset,\emptyset,\emptyset,\emptyset).
\]
We can now compute $E_1(D)$. All rules but $r_5$ are applicable 
since their antecedent is a subset of $\partial^+_0(D)$. Moreover, 
for $r_1$, $r_2$, $r_3$, $r_4$ and $r_7$, there are no rules for 
the complement of their conclusion, thus, vacuously, the condition 
that all rules for the opposite are either defeated or discarded is
satisfied. Hence we add $\neg a$ to $\partial^+_1(D)$, and 
$\partial^{+\OBL}=\set{a,b,d}$. Given that there are no constitutive rules
for $a, b,\neg b, c,\neg c, d, \neg d$ these literals are all in 
$\partial^-_1(D)$. For the same reason 
$\partial^{-\OBL}=\set{\neg a, \neg b,\neg d}$.  Notice that, even 
if we have an applicable prescriptive rule for $c$ ($r_6$), there is a 
prescriptive rule for $\neg c$ ($r_5$), but we are not able to 
assess, yet, whether it is applicable or discarded.  We 
are not in the position to populate $\partial^{+\wedge}_1$ since 
$\partial^{+\OBL}_0$. For $\partial^{-\OBL}_1$ we can compute the
reduct for all individual literals, and determine what literals are 
not in $\partial^{+\OBL}(\reduct(D,\set{l}))$. In the theory, the only 
constitutive rule is $r_4$, and we can repeat the argument for 
$\partial^{-\OBL}_1(D)$.  Accordingly, $\partial_1^{-\wedge}(D)$
contains all conjunctions where at least one element belongs to 
$\partial^{-\OBL}(D)$, e.g., $\neg a \wedge b$, 
$a\wedge \neg b\wedge c$ and so on.

We proceed to the computation of $E_2$, specifically 
$\partial_2^{+\wedge}(D)$. We have $a, b, d\in \partial_1^{+\OBL}(D)$.
Thus, we have to consider what combinations result in conjunctive
obligations that are not pragmatic oddity instances. To this end, 
we compute $\reduct(D,\set{\neg a})$, $\reduct(D,\set{\neg b})$, and 
$\reduct(D,\set{\neg c})$. Since there are no constitutive rules for 
$\neg b$ and $\neg c$,  
$\reduct(D,\set{\neg b})=\reduct(D,\set{\neg b})=D$, and we have seen
that $a,b,d\in\partial^{+\OBL}(D)$. Removing $\neg a$ results in 
$\neg a\in\partial^-(\reduct(D,\set{\neg a}))$ making $r_1$ not
applicable for $b$ at index 2. However, we still have rule $r_3$ to
include $b$ in $\partial^{+\OBL}(\reduct(D,\set{\neg a}))$. Hence, 
$a\wedge b\in\partial_2^{+\wedge}(D)$, and so are $a\wedge d$, 
$b\wedge d$ and $a\wedge b\wedge d$. 

For $E_3(D)$, we have two applicable prescriptive rules for complementary 
literals: $r_5$ for $\neg c$ and $r_6$ for $c$. However, we do not have
instances of the superiority relation for them. Thus, $c$ and $\neg c$
are not provable as obligations, and we include them in 
$\partial_3^{-\OBL}(D)$.   This, in turn, allows us to establish that 
$c\wedge d\in\partial_4^{-\wedge}(D)$. After this step we no longer add
 elements to the extension, meaning that we have reached the fixed point. 
\end{example}

\begin{example}
Let us turn our attention to the theory $D$ for the multiple dependencies
scenarios, where the rules are
\begin{align*}
	r_1\colon \To_\OBL a\phantom{a}        
	&& 
	r_2\colon {} \To_\OBL b \phantom{b}\\
	r_3\colon \neg a \To_\OBL c 
	&& 
	r_4\colon \neg b \To_\OBL c
\end{align*}	
where $F=\set{\neg a,\neg b}$.  It is easy to verify that 
$a,b,c\in\partial_1^{+\OBL}$. Let us consider the reducts for 
$\set{\neg a}$, $\set{\neg b}$ and $\set{\neg a, \neg b}$. For 
the first $F=\set{\neg b}$; therefore 
$\neg a\in\partial^-(\reduct(D,\set{\neg a})$ and $r_3$ is discarded.
However, we can still use $r_4$ to conclude 
$\OBL b$ ($b\in\partial^{+\OBL}(\reduct(D,\set{\neg a})$. Accordingly 
$a\wedge c\in\partial_2^{+\wedge}(D)$. We can repeat a similar argument 
for $\reduct(D,\set{\neg b})$ to determine that 
$b\wedge c\in \partial_2^{+\wedge}(D)$. Finally, for $a\wedge b\wedge c$
we notice that when we remove both $\neg a$ and $\neg b$ from the set 
of facts in the computation of $\reduct(D,\set{\neg a,\neg b})$, rules
$r_3$ and $r_4$ are both discarded, and there are no remaining 
prescriptive rules for $c$; ergo, $c\in\partial^{-\OBL}(\reduct(D,\set{\neg a,\neg b}))$, 
which implies $a\wedge b\wedge c\in\partial^{-\wedge}(D)$. 
\end{example}

\begin{definition}
An extension 
\[\langle
  \partial^+(D),
  \partial^-(D),
  \partial^{+\OBL}(D),
  \partial^{-\OBL}(D),
  \partial^{+\wedge}(D),
  \partial^{-\wedge}(D)
  \rangle\]
is \emph{coherent} if $\partial^+\cap\partial^-=\emptyset$, 
$\partial^{+\OBL}\cap\partial^{-\OBL}=\emptyset$ and 
$\partial^{+\wedge}\cap\partial^{-\wedge}=\emptyset$.
  
An extension is \emph{consistent} if for every set $\partial^*$, it is not 
the case that $p$ and $\non p$ are both in $\partial^*$.
\end{definition}
Intuitively, coherence says that no literal is simultaneously provable and unprovable. Consistency says that a literal and its negation are not both defeasibly provable. 
\begin{proposition}
  Given a theory $D$, $E(D)$ is coherent. If $F$ does not contain a pair of 
  complementary literals, and the transitive closure of $>$ is acyclic, then 
  $E(D)$ is consistent. 
\end{proposition} 
\begin{proof}
Notice that the conditions to establish that a literal/conjunction is a member of 
one of the positive sets of an extension at a given stage are de facto the strong 
negation of the condition to add the literal to the corresponding negative set. 
We have to replace $a(r)\subseteq \partial^+$ for  $\forall a\in A(r), +\partial a\in P(1..n)$,
and $a(r)\cap \partial^-$ for $\exists a\in A(r), -\partial a\in P(1..n)$. 
Hence, we can use the results of \cite{igpl09policy}, see also Proposition~\ref{prop:consistency}.
Here we show the key cases for coherence. For the cases of consistency, see the 
proof in \cite{igpl09policy}.

We prove the proposition for coherence by induction on the extension's construction stage. 
The inductive base, the case for $E_0(D)$, is trivial by the definition of $E_0$.

For the inductive base, let us assume that coherence holds up to the $n$-th extension, 
$E_n(D)$. By the monotonicity of the construction, if a  rule is applicable at a step $m<n$,
then the rule remains applicable at step $n$ (similarly for discarded).   For $\partial^+$ and 
$\partial^-$, the argument is as follows: for a literal $l$ to be in $\partial^+{n+1}$, 
there must be a rule $r$ that is applicable at $E_n(D)$: by the inductive hypothesis, and 
Definitions~\ref{definition:BodyApplicable-e}, \ref{definition:APPL+pO-e} and  
\ref{definition:DISC+pO+pP-e} no rule is at the same time applicable and discarded for one 
and the same literal at the same time. This means that, for the condition for $\partial^-$,
there is a rule $s$ that is applicable in $E_n(D)$,  but then there is a rule $t$
applicable for $l$ at $E_n(D)$ and $t>c$, but for $\partial^-$ $t$ should either be discarded
or not stronger than $s$. Contradiction. Thus $\partial^+_{n+1}$ and $\partial^-_{n+1}$ 
are disjoint.
 
For $\partial^{+\OBL}\cap\partial^{-\OBL}=\emptyset$, we remark in addition to what we have 
just proved, we have to consider conditions 2 of Definitions \ref{definition:APPL+pO-e}
and \ref{definition:DISC+pO+pP-e} to realise by the inductive hypothesis that 
no rule can satisfy the conditions in the two definitions.

Finally, for $\partial^{+\wedge}\cap\partial^{-\wedge}=\emptyset$, by the inductive hypothesis 
$\partial^{+\OBL}_n\cap\partial^{-\OBL}_n=\emptyset$, in addition, the extension of any 
theory is unique (being the least upper bound of a finite monotonically increasing 
construction), and the reducts we consider are subsets of the given theory (thus, the 
coherence property holds for them as well).
\end{proof}
An inconsistency is possible only when the theory we started with was 
inconsistent (either because the facts are inconsistent or because the superiority
relation induces a cycle in the superiority relations, meaning that a rule is at 
the same time stronger and weaker than another rule). 
Accordingly, defeasible inference for defeasible deontic logic for pragmatic oddity 
does not introduce inconsistency. A logic is coherent (consistent) if the 
meaning of each theory of the logic, when expressed as an extension, is 
coherent (consistent).

\section{Complexity}
\label{sec:complexity}

In this section, we study the computational complexity of 
the problem of computing whether a conjunctive obligation is derivable 
from a given defeasible theory. To this end, we adapt the 
algorithm proposed in \cite{jpl:permission} to compute the extension 
of a defeasible theory, where the computation of the extension is 
linear in the size of the theory.  The algorithm is based on a series 
of transformations that reduce the complexity of the theory by either 
removing elements from rules when some elements are provable, or 
removing rules when they become discarded (and so no longer able to 
produce positive conclusions).

The paper aims to determine when conjunctive obligations are either 
provable or discarded.  Accordingly, we have to extend the definition to 
account for conjunctive obligations. However, if we want to maintain a
feasible computational complexity, we have to limit the conjunctions we
consider: given a set of $n$ literals, the set of all possible 
non-logically equivalent conjunctions that the $n$ literals can form
contains $2^n$ conjunctions; hence, we cannot compute in polynomial time for 
such a set if any element is derivable or refuted by the theory.  However, 
we are going to show that for each individual conjunction, we can compute in
polynomial-time whether it is derivable or refuted. 

\begin{definition}\label{sec:conjunctive-extension}
  Given a defeasible theory $D$, the \emph{conjunctive extension} of the 
  theory is the tuple: 
 \[
\langle 
  \partial^+(D), \partial^-(D),\partial^+_{\OBL}(D),\partial^-_{\OBL}(D) ,
  \partial^+_\wedge(D), \partial^-_\wedge(D)
\rangle
 \] 
where $\partial^+(D)$, $\partial^-(D)$, $\partial^+_{\OBL}(D)$ and 
$\partial^-_{\OBL}(D)$ are as in Definition~\ref{def:extension} and
\begin{itemize}
  \item $\partial^+_\wedge(D)$ is the set of conjunctive obligations appearing 
    in $D$ (i.e., $c=\OBL(c_1\wedge\cdots\wedge c_n)$ and $\exists r\in R$ such 
    that $c\in A(r)$) that are defeasibly provable in $D$;
  \item $\partial^-_\wedge(D)$ is the set of conjunctive obligations appearing 
  in $D$  that are defeasibly refutable in $D$.
\end{itemize}
\end{definition}
The algorithm to determine the conjunctive extension of a theory is based on the
following data structure (for the full details, we refer the reader to \cite{jpl:permission}).
We create a list of the atoms appearing in the theory.
Every entry in the list of atoms has an array associated to it. The array has ten 
cells, where every cell contains pointers to rules depending on whether and how the 
atom appears in the rule. 
The first cell is where the atom appears in the head of a constitutive rule, 
the second where the negation of the atom appears in the head of a constitutive rule, 
the third where the atom appears in the head of a prescriptive rule, 
the fourth where the negation of atom appears in the head of a prescriptive rule,
the fifth where the atom appears in the body of a rule,
the sixth where the negation of the atom appears in the body of a rule,
the seventh where the atom appears as an obligation in the body of a rule,
the eighth where the negation of the atom appears as an obligation in the body of a rule,
the ninth where the atom appears as a negative obligation in the body of a rule, and 
the tenth where the negation of the atom appears as a negative obligation in the body of a rule.
In addition, we maintain a list of conjunctive obligations occurring in the theory, 
and for every conjunction, we associate it to the rules where it appears in the 
body.

The algorithm works as follows: at every round, we scan the list of atoms. For
every atom (excluding the entries for the conjunctions), we look if the atom
appears in the head of some rules. If it does not appear in any of the cells
for the heads, we can set the corresponding literals as refuted; and we can
remove rules from corresponding cells. So, for example, given an atom $p$, if
there are no prescriptive rules for $\neg p$; then we can conclude that the
theory proves $-\partial_{\OBL}\neg p$; accordingly, all rules where
$\neg\OBL\neg p$ occurs in the body are (body)-discarded, and we can remove them
from the data structure. Similarly, if there are no constitutive rules for $\neg
p$, then we can prove $-\partial\neg p$, and then (i) all the rules where it
appears in the body are body-discarded, but also (ii) for each rule $r$ in whose
head $p$ appears as an obligation, no elements following $p$ in $r$ can any
longer be derived using $r$, and such elements are removed from the appropriate
cells. 
If an atom appears in the head of a rule, we determine (i) if the body of the
rule is empty and (ii) for prescriptive rules, if the atom is the first element 
of the head. If this is the case, then the rule is applicable, and we check
if there are rules for the negation. If there are no rules for the negation, or 
the rules are weaker than applicable rules, then the atom/literal is provable 
with the suitable proof tag. Then we remove the atom/literal from the 
appropriate rules. 
We repeat the above steps until we can no longer obtain new conclusions.
When we are not able to derive new conclusions, we turn our attention to the list
of the conjunctive obligations, where we invoke the following (sub)algorithm for
every conjunction $c=(c_1\wedge\cdots\wedge c_n)$ in the list (where
$C=\set{\non c_i, 1\leq i\leq n})$ 

\begin{algorithm}
\begin{algorithmic}[1]
\For{$i\in 1..n$}
   \If{$c_i\in\partial^-_{\OBL}(D)$}
      \State{$c\in\partial^-_\wedge(D)$ remove all rules $r$ where $c\in A(R)$}
      \State{Exit}
    \EndIf
    \If{$c_i\in\partial^+_\OBL(D)$}
      \If{$\forall c_j\neg c_i, \non c_j\in\partial^+(D)$}
        \If{$c_i\in+\partial^+_\OBL(\reduct(D,C\setminus\set{\non c_i}))$}
          \State{$i:=i+1$}
        \Else{ $c\in\partial^-_\wedge(D)$ remove all rules $r$ where $c\in A(R)$}
          \State{Exit}
        \EndIf
        \If{$\exists c_j\neq c_i, \non c_i\in\partial^-(D)$}
          \State{$i:=i+1$}
        \EndIf
      \EndIf
    \EndIf
  \State{Exit}
\EndFor
\State{$c\in\partial^+_\wedge(D)$, remove $c$ from all rules $r$ where $c\in A(r)$}
\end{algorithmic}
\caption{Evaluate Conjunctive Obligation $c=c_1\wedge\dots\wedge c_n$}
\label{algo}
\end{algorithm}

For every conjunction, the algorithm iterates over the conjuncts. The conjunction is not provable if a conjunct 
is not provable as an obligation (lines 2--4). 
If the conjunct is provable as an obligation, it checks whether the 
violations of the other obligations are provable; if so, it has to check 
whether the obligation of the conjunct is independent of the violations. 
To determine this, we can repeat the whole algorithm with the sub-theory
obtained by the transformation $\reduct(D,C\setminus\set{c_i})$. If it is
independent, we continue with the next element of the conjunction; 
otherwise, the conjunction is not derivable. Similarly, if some 
of the violations are not derivable, we continue with the iteration. 
The conjunction is provable when the iteration is successful for all the 
conjunction elements. 

At the end of the sub-routine, we return to the main algorithm; if there are
changes in the rules, we repeat the process; otherwise, the process terminates. 
 
\begin{proposition}
The algorithm to compute the conjunctive extension of a theory $D$ computes
the extension $E(D)$ when the language is restricted to the conjunctive obligations
that occur in $D$.
\end{proposition}
\begin{proof}
The algorithm consists of two parts. The first part is the algorithm presented
in \cite{jpl:permission} to compute the extension of a Defeasible Deontic Logic. 
The proof conditions presented in this paper are restrictions of those in
\cite{jpl:permission}, and they are equivalent as far as the language in this 
paper is concerned. The language (and algorithm) in \cite{jpl:permission} does 
not allow for conjunctive obligations. Thus, we can consider each conjunctive 
obligation with a new literal. \cite{jpl:permission} proves that their 
algorithm is sound and complete to compute the extension (corresponding to 
$
\langle 
  \partial^+(D), \partial^-(D),\partial^+_{\OBL}(D),\partial^-_{\OBL}(D) 
\rangle
$). The second part of the computation presented in this paper is Algorithm~\ref{algo},
that effectively acts as an external oracle to determine whether the 
conjunctive obligations (the new literal) hold or not (based on the reduct 
construction). If a conjunctive 
obligation holds then it can be added to $\partial^+_\wedge(D)$, otherwise to
$\partial^-_\wedge(D)$, and we can resolve the corresponding new literal. 
Thus, the correctness depends on the correctness of Algorithm~\ref{algo}
against the construction in Definition~\ref{def:fix-point}. The explanation 
of the algorithm above shows that the steps in the algorithm correspond 
to the steps to compute $\mathcal{T}(\partial^{+\wedge}_{n}(D))$ and 
$\mathcal{T}(\partial^{-\wedge}_{n}(D))$.
\end{proof}

Concerning the computational complexity, \cite{jpl:permission} proves that the 
complexity of computing the extension of a defeasible theory without conjunctive 
obligations is linear in the size of the theory, where the size of the theory is 
determined by the number of symbols in the theory, and hence if $n$ and $r$ 
stand for, respectively, the number of atoms and the number of rules in the 
theory, the complexity is in $O(n*r)$.    
For the complexity of computing the conjunctive extension of a defeasible
theory, we have to take into account the complexity of the Evaluate Conjunctive 
Obligation algorithm and the number of times we have to compute it. This can be
determined as follows: let $m$ be the number of conjunctive obligations in the 
theory, and $k$ the number of conjuncts in the longest conjunctive obligation. 
For each of them, we have to compute the extension of $\reduct(D,C)$, thus we have 
to perform $O(m*k*O(n*r))$ computations on top of the calculation of the 
extension (i.e., $O((m+n)*r)$).

\begin{proposition}\label{prop:complexity}
The conjunctive extension of a theory can be computed in polynomial time. 
\end{proposition}   

Notice that the algorithm Evaluate Conjunctive Obligation can be used 
to evaluate any conjunctive obligation, not only the conjunctive obligations 
occurring in a theory. All we have to do is to compute the conjunctive 
extension of the theory and then evaluate the single conjunctive obligation, 
and as we have just seen, this can be calculated in polynomial time. 

\section{Summary and Discussion}
We have proposed an extension of Defeasible Deontic Logic that prevents
the so-called Pragmatic Oddity paradox from occurring. The mechanism we used to achieve 
this result was to provide a schema that allows us to give a guard to the 
derivation of conjunctive obligations ensuring that each individual obligation
does not depend on the violation of the other obligation. The proof theory of defeasible logic gives the mechanism; in addition, we presented a 
bottom-up characterisation of the logic that avoids the problem of
non-deterministically selected derivations.  Furthermore, the bottom-up
construction is the foundation of the algorithm presented in 
\cite{jpl:permission} to compute the extension of a defeasible 
deontic theory (without conjunctive obligations) in linear time. This allows us to give a
polynomial upper bound to the problem of computing the extension of a
defeasible theory with pragmatic oddity (limiting to the conjunctive 
obligations appearing explicitly in the theory). First, we treat the 
conjunctive obligations in a theory as new literals, and then for each of 
them, we spin out the computation of the extensions for the relevant reducts. 
While the upper bound complexity of the logic is polynomial and hence 
feasible, the algorithm we just outlined is not optimal.  
Most practical real-life examples are likely to involve only a few 
conjunctive obligations, and ones with few conjuncts, so modest 
inefficiency of the algorithm for implementation is often not a serious
practical problem. Nonetheless, it is desirable, as a next step, to devise an 
optimal algorithm to implement these novel proof conditions and the  
bottom-up procedure.

\subsection*{Acknowledgments}
Preliminary versions of the paper proposing the idea of the logic were
presented at Jurix 2019 \cite{Governatori:2019:A-Computational} and 
DEON 2020/2021 \cite{deon20:pragmatic}.
We thank the anonymous reviewers for their valuable comments on earlier 
versions of the paper. 

\bibliographystyle{plain}
\bibliography{biblio}
\end{document}